\documentclass[%
reprint,
amsmath,amssymb,
aps,
floatfix,
]{revtex4-2}

\usepackage{graphicx}
\usepackage{dcolumn}
\usepackage{bm}
\usepackage{hyperref}
\usepackage{textcomp}%
\usepackage{wrapfig}
\usepackage{ragged2e}

\begin{document}
\hyphenpenalty=1000
\exhyphenpenalty=1000
\clubpenalty=0
\widowpenalty=0


\title{Study of vacancy ordering and the boson peak in metastable cubic Ge-Sb-Te using machine learning potentials}

\author{Young-Jae Choi}
\author{Minjae Ghim}%
\author{Seung-Hoon Jhi}
 \email{jhish@postech.ac.kr}
\affiliation{
Department of Physics, Pohang University of Science and Technology, Pohang 37673, Republic of Korea
}




\date{\today}

\begin{abstract}
\raggedright
\begin{wrapfigure}{r}{0.45\textwidth}
  \centering
  \includegraphics[width=0.45\textwidth]{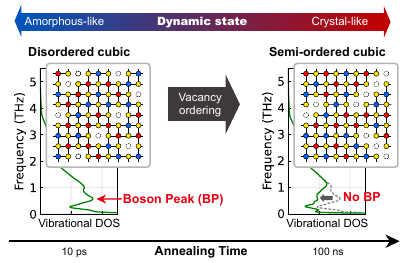}
\end{wrapfigure}
\justifying{}
~~~The mechanism of the vacancy ordering in metastable cubic Ge-Sb-Te \mbox{(c-GST)} that underlies the ultrafast phase-change dynamics and prominent thermoelectric properties remains elusive.
Achieving a comprehensive understanding of the vacancy-ordering process at an atomic level is challenging because of enormous computational demands required to simulate disordered structures on large temporal and spatial scales.
In this study, we investigate the vacancy ordering in \mbox{c-GST} by performing large-scale molecular dynamics simulations using machine learning potentials.
The initial \mbox{c-GST} structure with randomly distributed vacancies rearranges to develop a semi-ordered cubic structure with layer-like ordered vacancies after annealing at 700~K for 100~ns.
The vacancy ordering significantly affects the lattice dynamical properties of \mbox{c-GST}.
In the initial structure with fully disordered vacancies, we observe a boson peak, usually associated with amorphous solids, that consists of localized modes at $\sim$0.575~THz.
The boson peak modes are highly localized around specific atomic arrangements of straight vacancy-Te-vacancy trios.
As vacancies become ordered, the boson peak disappears, and the Debye-Waller thermal \textit{B} factor of Te decreases substantially.
This finding indicates that the \mbox{c-GST} undergoes a transition from amorphous-like to crystalline-like solid state by thermal annealing in low-frequency dynamics.
\end{abstract}

\maketitle


\section{\label{sec:1}Introduction}
Chalcogen-based phase-change materials are considered promising candidates for next-generation non-volatile memory devices~\cite{RN39, RN61, RN40}.
Ge-Sb-Te (GST) ternary compounds especially have attracted great attention due to the high stability of their amorphous solid phase and the convenient phase-switching operations by either electric, thermal, or laser pulses~\cite{RN61}.
While GeTe exhibits a drastic change in resistivity ($\sim$$10^6$ fold) by the amorphous to crystalline phase transition at \textit{T}$\sim$450~K, crystalline GST displays a multilevel resistivity spectrum over a 1000-fold range when annealed at 400~K$<$\textit{T}$<$600~K~\cite{RN42}.
Amorphous GST (\mbox{a-GST}) crystallizes into a metastable rock-salt cubic phase \mbox{(c-GST)} before further crystallizing into a hexagonal phase~(\mbox{h-GST}).
In \mbox{c-GST} lattice structures, anions~(Te) occupy 4a sublattice sites, while cations~(Ge/Sb) and vacancies occupy 4b sites~\cite{RN25}.
Stable \mbox{h-GST}, obtainable by high-temperature annealing at \textit{T}$>$500~K, has a layered structure with alternating stacks of anion and cation layers~\cite{RN42}.
Several types of hexagonal phases are distinguished by the sequence of layer stacking, with the most stable and commonly observed being the Kooi sequence~(vacancy-Te-Sb-Te-Ge-Te-Ge-Te-Sb-Te)~\cite{RN43}.

GST phase-change memory devices exhibit fast amorphous to crystalline phase switching by electric pulses, which occur within 100~ns~\cite{RN40}.
A few models have been proposed as a microscopic origin of the rapid crystallization of GST.
Kolobov \textit{et al.} proposed that the rapid phase transition of GST is possible by the facile transfer of cations from a tetrahedron center in its amorphous state to an octahedron center in its cubic state, namely the umbrella-flip distortion model~\cite{RN26}.
Hegedüs and Elliott conducted \textit{ab-initio} molecular dynamics~(AIMD) simulations of the amorphous-to-cubic phase transition in a 63-atom Ge$_2$Sb$_2$Te$_5$~(GST225) system and found that the fast crystallization of \mbox{a-GST} was caused by the excessive formation of 4-fold rings by melt-quenching~\cite{RN47}.
The growth rate of a GST crystal domain is also fast, ranging 10$^{-9}$$\sim$3 m/s, depending on the annealing temperature~\cite{RN63, RN41}.
Lee and Elliott~\cite{RN17} demonstrated, through an AIMD annealing simulation of a 180-atom \mbox{a-GST} system, that vacancies located near amorphous-crystal interfaces may promote the rapid growth of a crystal domain.
Recently, the change in electronic structure by crystallization was investigated using the maximally localized Wannier function method from a large-scale AIMD simulation of a 328-atom system~\cite{RN28}.
The complex configuration of Wannier electrons in \mbox{a-GST}, including many weak bonding states and lone-pair electrons, is advantageous for the fast structural transition accompanied by frequent bond-breaking and formation processes~\cite{RN28}.

The \mbox{a-GST} structures build \textit{s}-\textit{p} hybridized covalent bonding states, while well-ordered crystalline structures build directional \textit{p}-orbital resonant bonding states, which leads to a stark optical contrast between the two structures~\cite{RN36}.
The directional \mbox{\textit{p}-bonding} could become unstable against $sp^3$ bonding as the \mbox{\textit{p}-bonding} chain length increases~\cite{RN7}.
A signature of the \mbox{\textit{p}-bonding} instability is the softening of specific phonon modes, resulting in the distortion observed in $\alpha$-GeTe with alternating long and short bonds~\cite{RN6}.
In this sense, the presence of vacancies can stabilize the \mbox{\textit{p}-bonding} chains, and their distribution is a key for understanding the structural and electronic properties of \mbox{c-GST}.

The distribution of vacancies in \mbox{c-GST} can be either random or ordered to a certain degree~\cite{RN34}.
The vacancy-ordered-cubic~(VOC) structures slightly differ from the hexagonal structures in terms of interlayer distance and stacking geometry of Te-Te van der Waals~(vdW) layers~\cite{RN34}.
A typical hexagonal structure obtained by annealing is the Matsunaga phase, which is similar to the Kooi phase but with numerous substitutional defects of Ge and Sb atoms~\cite{RN29}.
A major structural change in the cubic-to-hexagonal phase transition is thus accompanied by vacancy ordering.
The electrical resistance of crystalline GST, which varies over a 1000-fold range, reflects vacancy ordering and exhibits a feature of the metal-insulator transition at a certain threshold~\cite{RN42}.
Additionally, GST-based phase-change memories show a reduction in crystallization temperature and electrical resistance with an increase in operation cycles~\cite{RN41}.
A non-Arrhenius crystallization of \mbox{a-GST} was observed, as the activation energy underwent a discontinuous 6-fold magnitude change at \textit{T}$\sim$480~K~\cite{RN41}.
All these observations in phase-change memories are likely attributed to the distribution of vacancies and their ordering.
However, they cannot be tracked atomistically by measurements nor by first-principles simulations due to their complexity in length and time scales.

The atomic-scale studies of GST on electronic and structural properties require a substantial size in atomic number and simulation time.
Direct AIMD simulations are limited to $\sim$1000 atoms due to high computational costs~\cite{RN18, RN51, RN52, RN53, RN54}.
Recently, machine learning potential~(MLP) methods have expanded the simulation capacity to an unreachable scale by AIMD, allowing atomistic calculations with an accuracy of first-principles methods~\cite{RN11, RN14, RN16, RN12}.
The medium-range order of large-sized ring configurations has been identified and the first sharp diffraction peak of X-ray structure factor in measurement has been reproduced~\cite{RN11}.
The mid-gap states in amorphous GST225~\cite{RN14} and quench-rate-dependent crystallization~\cite{RN16} have also been directly addressed in machine-learning-potential molecular dynamics~(MLMD) without resort to phenomenological modeling.
In this study, we used MLMD methods to simulate the thermal annealing of \mbox{c-GST} at 700~K and analyzed the distribution and rearrangement of vacancies.

\section{\label{sec:2} Methods}
\subsection{\label{sec:2-1}\texorpdfstring{Ge$_2$Sb$_2$Te$_5$}{Ge2Sb2Te5} potential}
We utilized the Neural E(3)-equivariant interatomic potentials~(NequIP) model to develop MLPs for GST225~\cite{RN3}.
NequIP is based on graph neural networks and learns a high-dimensional potential energy surface by partitioning the potential energy of a system into individual atomic contributions, namely atomic energies~\cite{RN3}.
The NequIP process involves multiple iterations of message passing to effectively convey non-local structural information~\cite{RN3}.
The message passing operates with feature tensors that contain structural information and are equivariant under rotation, reflection, and translation transformations~\cite{RN3}.

The NequIP potential was constructed with a local cutoff radius $r_c$ of 4~\AA{}, including 7.40 atoms on average inside a local environment sphere.
The maximum rotation order $l_\textrm{max}$ of the feature tensors was set to be 3, as the potentials with $l_\textrm{max}$=3 reproduce the angular distribution function of liquid GST225 at 1100~K better than those with $l_\textrm{max}$=2~(see Supplemental Material for further information on the NequIP training parameters~\cite{RN35}).
The converged GST225 potentials yielded an energy root-mean-square error~(RMSE) of 51.8~meV/atom and a force RMSE of 74.7~meV/\AA{} for the whole validation dataset.
We note that the RMSE values can vary consequentially for each class of the training dataset.
For the \mbox{c-GST} dataset, which is the primarily focused phase of this study, the RMSEs are 33.8~meV/atom for energies and 43.8~meV/\AA{} for forces~(see Supplemental Material for the validation of the GST225 potential~\cite{RN35}).
Inclusion of a dataset in a wider energy window inevitably increases the RMSEs but covers the potential energy surface more broadly, ensuring more vigorous stability of molecular dynamics simulations~\cite{RN4}.

The resulting GST225 potentials simulated properly the melting and crystallization processes (see Supplemental Material for the potential energy change during melting and crystallization~\cite{RN35}).
The melting point of \mbox{c-GST} in our simulation was found to be $\sim$1000 K, which is comparable to the measurement of $\sim$900 K~\cite{RN142}.
A slight discrepancy may be attributed to some factors neglected in our simulations such as surfaces or structural defects that accelerate the melting process.
The amorphous-to-cubic crystallization time was estimated to be about 400$\sim$700 ps, which is similar to the results in previous studies (400$\sim$600 ps) using a different type of MLP~\cite{RN11}.

\subsection{\label{sec:2-2}Randomized Atomic-system Generator (RAG)}
To develop the GST225 potentials, we constructed a training dataset using the randomized atomic-system generator~(RAG) method~\cite{RN4}.
The RAG is a novel approach for creating a quality training set for MLPs~\cite{RN4}.
While AIMD simulations provide reliable data, they are limited in covering the diverse atomic configurations and thermodynamic ensembles of the potential energy surfaces~\cite{RN4}.
In contrast, simple random sampling probes a wide range of configurations but overlooks local minima with unphysical over-weighting on high-energy configurations~\cite{RN4}.
The RAG method overcomes these limitations by utilizing the advantages of random sampling and structural optimization from both approaches.

The RAG method generates the initial structures from randomized lattice structures, species arrangements, and atomic displacements.
These initial structures from the RAG sampling are then optimized through 10$\sim$15 iterative steps of force minimization by DFT calculations, resulting in a training set that encompasses a wide spectrum of potential energy states.
The resultant RAG dataset covers a broad potential surface, including local minima, which are likely to be missed by AIMD or simple random sampling.
Since the RAG scheme constructs efficiently a diverse and comprehensive training set for MLPs, it is particularly useful for complex systems such as liquids, amorphous solids, and defective crystals with less computational loads than AIMD construction.

The GST225 RAG dataset comprised 66457 atomic configurations, covering a broad range of structural phases (liquid, amorphous solid, metastable cubic, and hexagonal) with potential energies ranging \mbox{--4.2} $\sim$ \mbox{--3}~eV/atom (see Supplemental Material for further information on the RAG dataset and its potential energy plot~\cite{RN35}).
Approximately 5\%~(3457 configurations) of the RAG dataset was randomly selected to form the validation set, and the remaining $\sim$95\%~(63000 configurations) constituted the training set.

Our RAG training dataset encompasses an extensive array of configurations, including single elements (Ge, Sb, Te), binaries (GeTe, Sb2Te3), and ternaries, along with a variety of segregated structures, covering a diverse range of stoichiometries and structures.
This breadth is validated through t-SNE analysis~\cite{RN4}, which visualizes data distribution by reducing non-primal dimensionalities using the similarity check~\cite{RN121}.
Extrapolation capability of MLPs across various stoichiometries was also demonstrated in a recent study of GST compounds ~\cite{RN145}.

\subsection{\label{sec:2-3}Linear chain of \textit{p}-bonding}
We employed the linear chain of \mbox{\textit{p}-bonding}~(\mbox{\textit{p}-bonding} chain) model to analyze the structure of \mbox{c-GST}~\cite{RN7}.
From a stoichiometric perspective, the \mbox{\textit{p}-bonding} network in GeTe is simple, with a formula unit of GeTe possessing six valence \textit{p}-electrons and their $p_x$-, $p_y$-, and $p_z$-orbitals tending to align in a linear chain-like structure.
In contrast, Sb$_2$Te$_3$ features a vdW-gapped structure with successive stacking of quintuple layers, containing 18 \textit{p}-electrons in a formula unit.
This stoichiometric interpretation applies to various chemical compositions of the pseudo-binary compounds~(GeTe)$_m$--(Sb$_2$Te$_3$)$_n$.

The \mbox{\textit{p}-bonding} chain of \mbox{c-GST} in this study was defined as the sequential bonding of cations and anions with restricted bonding connections that are less than 4~\AA{} in length and larger than 135\textdegree{} in dihedral angle.
The first peak of the radial distribution function~(RDF) of \mbox{c-GST} at 700~K terminates at 4~\AA{}~(see Supplemental Material for the RDF and ADF of \mbox{c-GST} at 700~K~\cite{RN35}).
Every linear \mbox{\textit{p}-bonding} chain in \mbox{c-GST} is truncated by two vacancies, and its length is measured by the number of \mbox{\textit{p}-bonding} units or equivalently by subtracting one from the number of constituent atoms.
Notably, the structural network of \mbox{\textit{p}-bonding} chains is fully determined by vacancy distribution, and thus its distribution can serve as a gauge of the vacancy ordering.

\subsection{\label{sec:2-4}Vibrational density of states~(VDOS)}
The dynamical state of a simple harmonic oscillator can exhibit autocorrelation.
Even when the dynamical phase of an oscillator stochastically collapses due to its finite lifetime, autocorrelation can still be picked for a large ensemble.
The density of the vibrational states of the oscillators can be calculated by Fourier transforming the velocity autocorrelation functions,
\begin{eqnarray}
g (\omega)= \frac{1}{N} \sum^{N}_{j}{\int^{\infty}_{-\infty}{dt\ e^{i\omega t} \gamma_j (t)}}
\label{eq:1}\;,\\
\gamma_j (t) = \frac{\langle \textbf{v}_j(t+\tau)\cdot \textbf{v}_j(\tau) \rangle_\tau}{\langle |\textbf{v}_j (\tau)|^2 \rangle_\tau}
\label{eq:2}\;,
\end{eqnarray}
where $g(\omega)$ is the vibrational density of states~(VDOS) for \textit{N} atoms, $\gamma_j(t)$ the atomic velocity autocorrelation function~(VACF), $v_j(t)$ the velocity vector of atom \textit{j}, and $\langle f(\tau) \rangle_\tau$ represents the ensemble average over $\tau$~\cite{RN37, RN38}.
A long~($>$10~ps) and large~($>$1000 atoms) simulation is necessary to obtain a reliably converged VDOS.
The frequency shift and linewidth broadening of phonons caused by their anharmonicity can be effectively captured by classical MD simulations.
This method provides a reliable distribution of vibrational modes at temperatures above the Debye temperature $T_{\mathrm{D}}$, as guaranteed by the energy equipartition theorem~\cite{RN37}.
We note that the VACF captures not only phonons but also localized atomic vibration modes.

The VDOS of typical crystalline solids is proportional to $\omega^2$ at low frequencies according to the Debye model.
The reduced VDOS, $g(\omega)/\omega^2$, is used to investigate the low-frequency dynamics that determines the thermal transport properties.
For perfect crystals, no peaks appear in the reduced VDOS below the frequency of a Brillouin peak, which originates from Van Hove singularity by the transverse acoustic~(TA) phonons near the Brillouin zone boundaries~\cite{RN140}.
In contrast to the Brillouin peaks, there may exist another kind of peak in the VDOS spectrum below the Brillouin peaks in amorphous solids.
These peaks are called boson peaks and are known to originate from structural disorder~\cite{RN140, RN31}, low mechanical rigidity~\cite{RN141}, or low atomic density~\cite{RN139}.

\subsection{\label{sec:2-5}DFT calculations}
The potential energies and forces of the RAG configurations were calculated using the Vienna \textit{Ab initio} Simulation Package~(VASP) with the projector augmented-wave~(PAW) method~\cite{RN56, RN57}.
The Perdew-Burke-Ernzerhof functional was adopted for the exchange-correlation term~\cite{RN58}, and the DFT-D3 method with Becke-Johnson damping function was used to account for additional van der Waals energy correction~\cite{RN65, RN66}.
The cutoff energy for the plane-wave basis set was set to 520 eV, and momentum space sampling was performed with an isotropic spacing of 0.13~\AA{}$^{-1}$.
We note that structural parameters can be affected by the choice of pseudopotentials~\cite{RN23, RN144, RN50}.
In particular, it has a long history that electronic and structural properties are sensitive to the way that semicore \textit{d} orbitals are treated in pseudizing as valence or core states.
Also, the type of exchange-correlation functional and treatment of long-range dispersion interaction may affect the total energy and the forces, and resultingly the training data set~\cite{RN11, RN143}.
Thus, care must be taken to construct the training data set that represent structural features properly.
Otherwise, unphysical consequences such as wrong bonds, under- or over-estimation of structural orders may result in. 
The atomic configurations were visualized using the Atomic Simulation Environment~(ASE) package~\cite{RN67}.

\section{\label{sec:3}Results}
\subsection{\label{sec:3-1}Vacancy ordering during annealing}

\begin{figure}[b]
\includegraphics[width=1.0\linewidth]{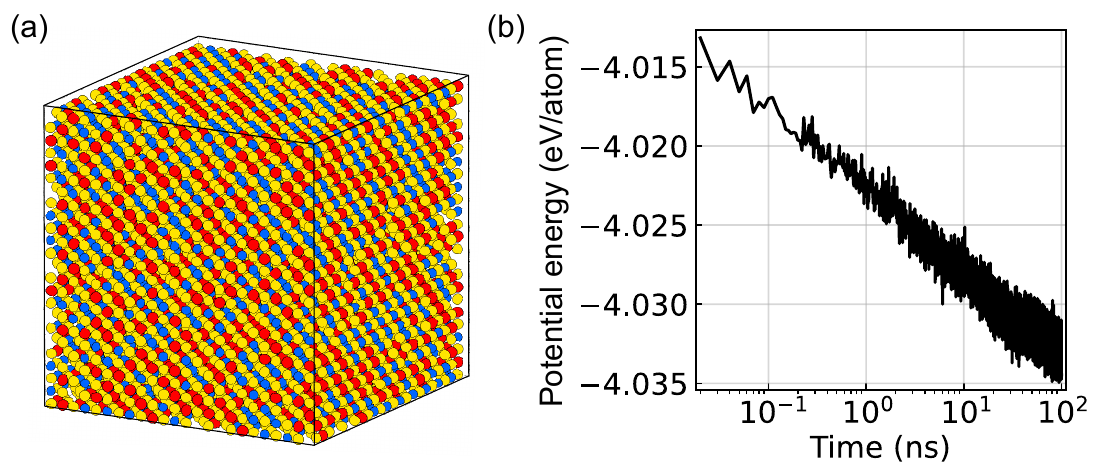}
\caption{\label{fig:1}
(a)~The initial RAG configuration of \mbox{c-GST} and (b)~the potential energy change during the annealing at 700~K.
The balls in blue, red, and yellow in~(a) correspond to Ge, Sb, and Te atoms, respectively.}
\end{figure}

The annealing of \mbox{c-GST} was simulated to investigate the rearrangement of cations and vacancies.
The initial atomic configuration in the annealing simulation was randomly generated by RAG~(Fig.~\ref{fig:1}a).
Our GST225 potentials reproduced successfully the \mbox{c-GST} structure through direct crystallization simulation from a melt-quenched amorphous structure.
The resulting structure was polycrystalline, exhibiting many domain walls.
To focus on the vacancy ordering analysis, we chose the RAG-generated structure as an initial configuration for the \mbox{c-GST} annealing simulation.
We verified that the RAG configuration contains local structures similar to the directly crystallized configuration and that the main conclusion in our simulations is not affected by the choice of initial \mbox{c-GST} configurations (see Supplemental Material for the crystallization of melt-quenched \mbox{a-GST}, the comparison of the \mbox{c-GST} structures constructed by RAG and direct crystallization simulation, and the structure and VDOS of \mbox{c-GST} directly crystallized from \mbox{a-GST}~\cite{RN35}).
The initial RAG configuration was the 12×12×12 supercell of the rock-salt FCC conventional cell, with 13824 lattice points in total.
All 6912 anion sublattices were occupied by Te atoms, while the remaining 6912 cation sublattices were occupied by 2765 Ge, 2765 Sb, and 1382 vacancies at random.
The cubic box was $\sim$72.17~\AA{} in size and had a mass density consistent with the experimentally measured value of \mbox{c-GST} $\sim$6.27~g/cm$^3$, or equivalently $\sim$0.0331~atom/\AA{}$^3$~\cite{RN64}.
The annealing simulation was carried out using the isothermal-isochoric~(NVT) ensemble at 700~K, at which the crystal growth rate is reported to be the highest~\cite{RN63}.
After a 100~ns annealing simulation, the \mbox{c-GST} system was transformed into a vacancy semi-ordered cubic state with the potential energy lowered by 20~meV/atom~(Fig.~\ref{fig:1}b).

\begin{figure}[t]
\includegraphics[width=1.0\linewidth]{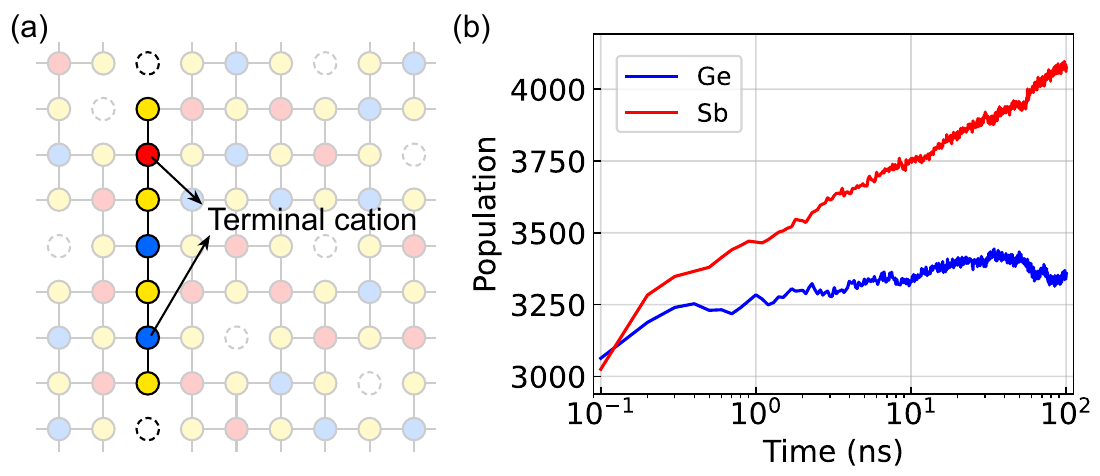}
\caption{\label{fig:2}
(a)~Schematic illustration of the terminal cation and (b)~changes in terminal-cation-site population.
In~(a), colored circles represent atoms and dashed circles denote vacancies.
The set of highlighted atoms in~(a) comprises the \mbox{\textit{p}-bonding} chain of length~6 truncated by two vacancies.
The population in~(b) refers to the number of atoms occupying the terminal-cation sites~(Ge: blue, Sb: red) as a function of the annealing time on a logarithmic scale~(all time points are shifted by 100~ps).}
\end{figure}

\begin{figure}
\includegraphics[width=1.0\linewidth]{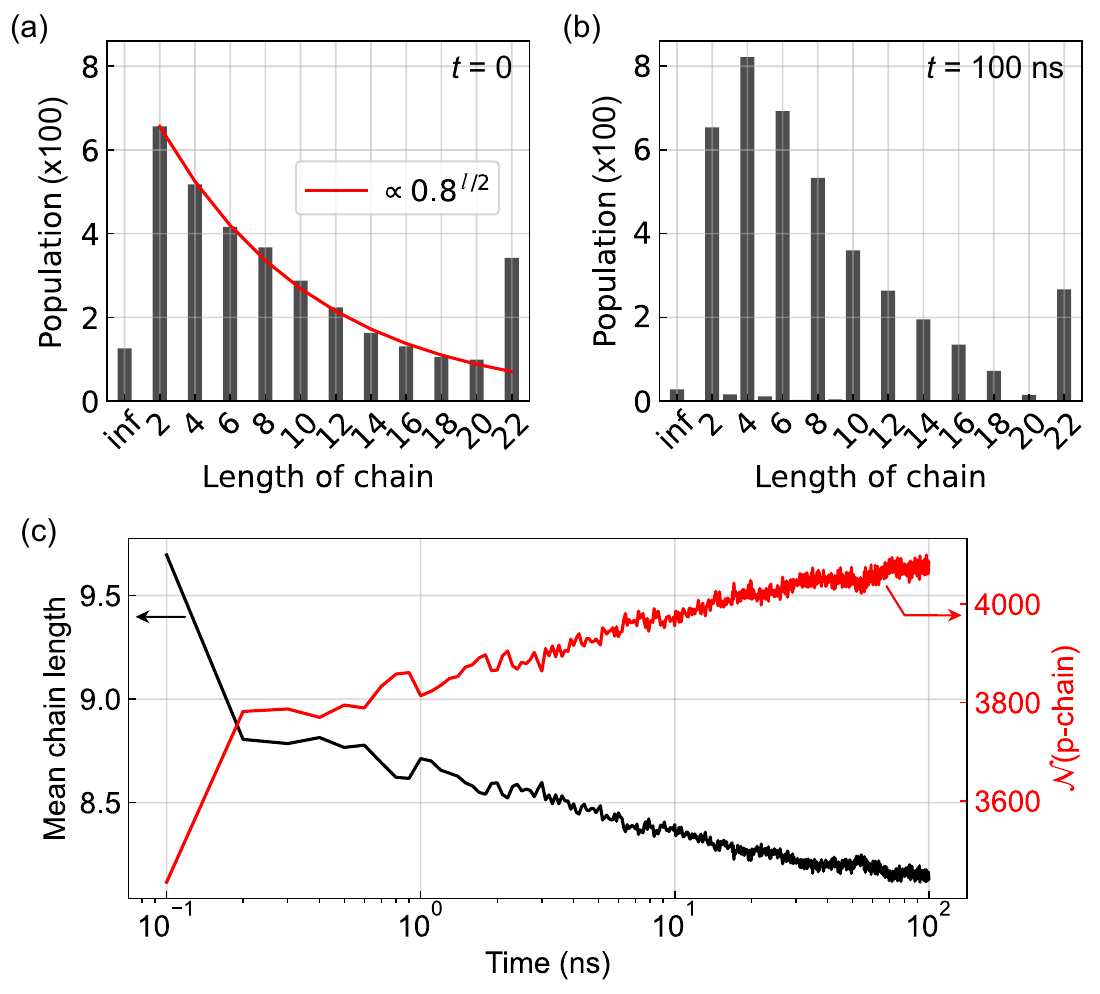}
\caption{\label{fig:3}
The \mbox{\textit{p}-bonding} chain length histogram at (a)~\textit{t}=0 and (b)~100~ns.
The \textit{x}-tick label “inf” indicates an infinitely long \mbox{\textit{p}-bonding} chain due to the open boundary condition, i.e., a \mbox{\textit{p}-bonding} chain of length~24 with the chain ends interconnected over the cell.
The fitting cu\textit{}rve in red in~(a) is $\sim$0.8$^{l/2}$ with \textit{l} as the chain length.
The sum of all \mbox{\textit{p}-bonding} chain lengths remains almost constant in the range of 33100$\sim$33300.
(c)~The change in the mean \mbox{\textit{p}-bonding} chain length~(black curve) and the number of \mbox{\textit{p}-bonding} chains~(red curve).}
\end{figure}

The changes in atomic structures were not noticed from RDF and ADF comparison (see Supplemental Material for the comparison of partial RDFs and total ADFs of \mbox{c-GST} annealed for 10~ps and 100~ns~\cite{RN35}). 
In order to investigate the structural changes caused by annealing, we monitored the population of cations at the terminal-cation sites of the \mbox{\textit{p}-bonding} chains~(Fig.~\ref{fig:2}).
The terminal-cation sites were occupied by Ge and Sb atoms at the same rate in the initial RAG configuration, but Sb atoms gradually outnumbered Ge atoms by annealing~(Fig.~\ref{fig:2}b).
After 100~ns of annealing, the number of Ge and Sb atoms occupying the terminal-cation sites were 3357 and 4073, respectively, which is an excess population of Sb by $\sim$21\%.
The formation of Sb-Te-vacancy configurations is a key structural change toward the hexagonal Kooi phase, in which \mbox{\textit{p}-bonding} chains always have Sb atoms at their terminal-cation sites.
As the vacancies became ordered, the number of \mbox{\textit{p}-bonding} chains increased, as did the population of Ge and Sb atoms at terminal-cation sites~(Fig.~\ref{fig:2}b and Fig.~\ref{fig:3}c).
We point out that a \mbox{\textit{p}-bonding} chain of length~2 has only one terminal-cation site.

The change in the number and length distribution of the \mbox{\textit{p}-bonding} chains during annealing was investigated~(Fig.~\ref{fig:3}).
In an infinitely large virtual \mbox{c-GST} with randomly distributed cations and vacancies, the length distribution of the \mbox{\textit{p}-bonding} chains converges to \mbox{$\textrm{P}(l)=x(1-x)^{l/2}$}, where \textit{x} is the ratio of vacancies to cation sites, and \textit{l} is the length of the \mbox{\textit{p}-bonding} chain.
For cubic GST225, \textit{x}=0.2 and the distribution of \mbox{\textit{p}-bonding} chain lengths follows the power law $\textrm{P}(l) \propto 0.8^{l/2}$.
The initial RAG configuration of a finite size with periodic boundary conditions shows a good fit to the power law distribution of \mbox{\textit{p}-bonding} chain lengths~(Fig.~\ref{fig:3}a).
After annealing, the number of \mbox{\textit{p}-bonding} chains with lengths $>$16 decreased, while the number of the \mbox{\textit{p}-bonding} chains with lengths 4$\sim$14 increased~(Fig.~\ref{fig:3}b).
The total number of the \mbox{\textit{p}-bonding} chains increased while the average length of the \mbox{\textit{p}-bonding} chains decreased to 8.14 by annealing~(Fig.~\ref{fig:3}c).
The vacancy-ordered configurations in the VOC and hexagonal Kooi phases of GST225 consist only of \mbox{\textit{p}-bonding} chains of length~8.

We defined the position of vacancies to investigate their distribution and migration.
Since vacancies in \mbox{c-GST} are not naturally identified, we defined their positions through the following steps.
First, we created an excessive number of vacancy candidates at the 3~\AA{}-prolonged positions along all the \mbox{\textit{p}-bonding} chains.
These candidates were then grouped into thousands of sets, following the rule that two vacancy candidates belong to the same set if their inter-vacancy-candidate distance is less than 2.2~\AA{}.
By regarding the candidates within a set as duplicates of the genuine vacancy, we determined the vacancy position as the mean position of all candidates within the set.
Faulty vacancies that had any neighboring atoms closer than 2~\AA{} were removed.
We note that defining a vacancy and its position in practical \mbox{c-GST} structures, especially at elevated temperatures, becomes ambiguous.
Despite the number of vacancies in the ideal \mbox{c-GST} structure being 1382, lattice distortions and atomic migrations in the actual annealing simulation at 700~K resulted in fewer vacancies: 1382, 1252, 1315, 1363, and 1373 vacancies were found at \textit{t}=0 s, 10~ps, 100~ps, 1~ns, 10~ns, and 100~ns, respectively.
See Supplemental Material for an example of the vacancy structure of \mbox{c-GST}~\cite{RN35}.

\begin{figure}
\includegraphics[width=1.0\linewidth]{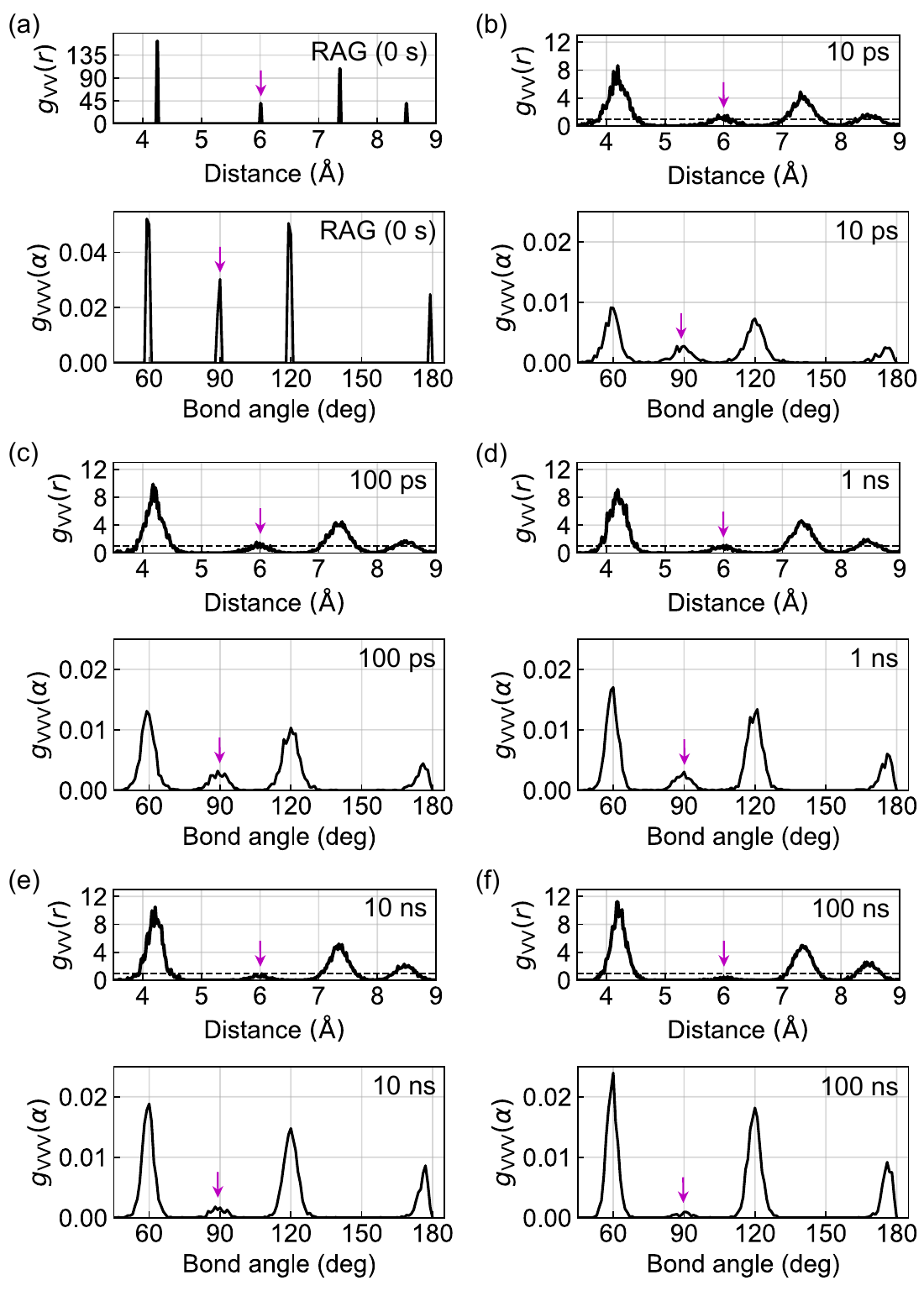}
\caption{\label{fig:4} 
Vacancy radial distribution function~($g_{\textrm{VV}}$) and vacancy angular distribution function~($g_{\textrm{VVV}}$) for (a)~0~s, (b)~10~ps, (c)~100~ps, (d)~1~ns, (e)~10~ns, and (f)~100~ns configurations, respectively.
The purple arrows indicate the peaks that should disappear in the vacancy fully ordered configurations.}
\end{figure}

The vacancy distribution was investigated using RDF and ADF~(angular distribution function) of vacancies~(Fig.~\ref{fig:4}).
The cutoff distance for the vacancy ADF calculation was 5.2~\AA{}, which encompasses the first peak of the vacancy RDF.
In the initial RAG configuration without any atomic displacement from the ideal \mbox{c-GST} lattice points, the sharp peaks were observed in the vacancy RDF at $\sim$3$\sqrt{2}$~\AA{}, 6~\AA{}, 3$\sqrt{6}$~\AA{}, and 6$\sqrt{2}$~\AA{}, and in the vacancy ADF at $\sim$60\textdegree{}, 90\textdegree{}, 120\textdegree{}, and 180\textdegree{}~(Fig.~\ref{fig:4}a).
Among the vacancy-RDF and ADF peaks, the 6~\AA{} and 90\textdegree{} peaks cannot exist in the vacancy fully ordered configurations of the VOC and hexagonal phases.
Throughout the annealing process, the intensities of the 6~\AA{} and 90\textdegree{} peaks were gradually reduced by the rearrangement of vacancies~(Fig.~\ref{fig:4}).
This reduction in the intensities of these two peaks is evidence of local vacancy ordering in the layers along the (111) direction.

\begin{figure}
\includegraphics[width=1.0\linewidth]{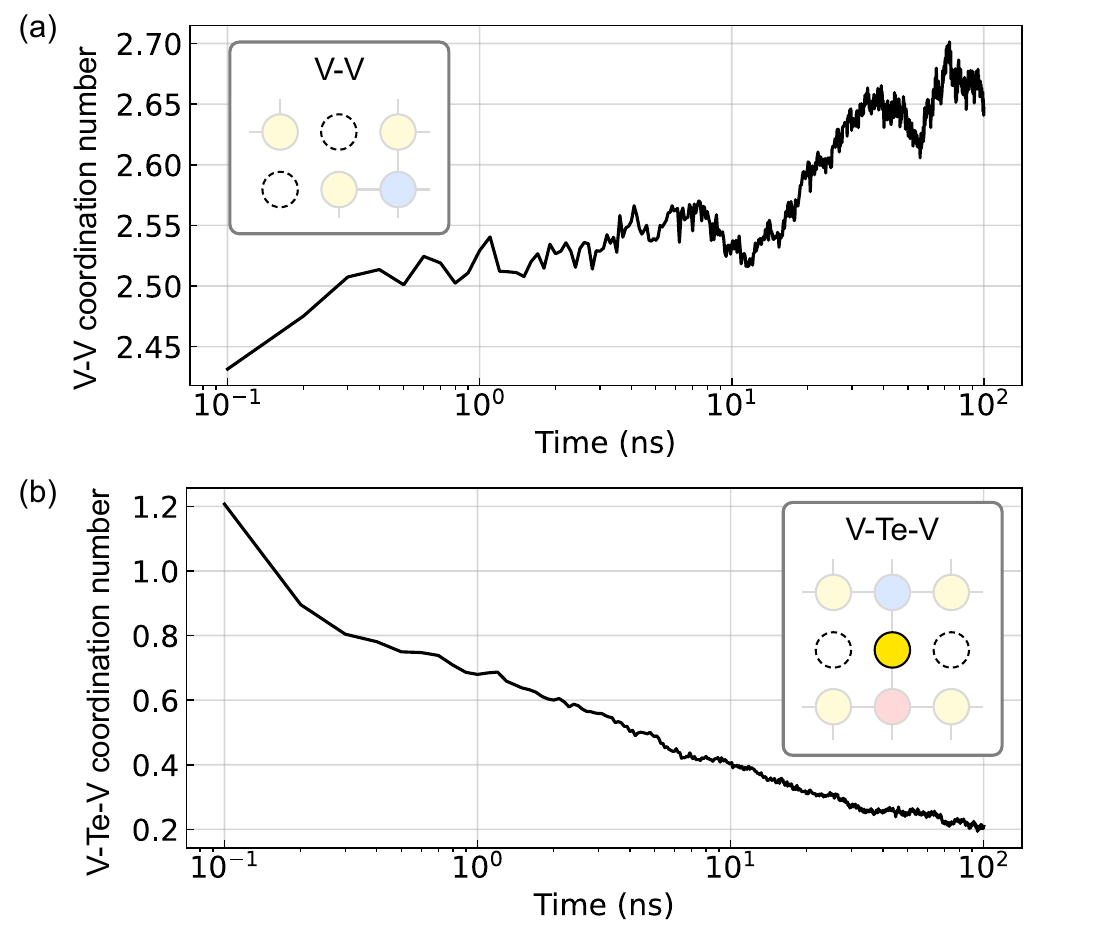}
\caption{\label{fig:5}
Change in the number of (a)~V-V duos and (b)~straight V-Te-V trios during annealing.
The insets illustrate example of the V-V duo and V-Te-V trio configurations.
The annealing time is given on a logarithmic scale shifted by 100~ps.}
\end{figure}

To analyze the increase in the vacancy-RDF first peak intensity at 3$\sqrt{2}$~\AA{} and the reduction of the vacancy-RDF second peak intensity at 6~\AA{} in a more specific way, the peak areas were calculated.
The mean coordination number of vacancy-vacancy (V-V) duos can be calculated by integrating the vacancy-RDF first peak~(Fig.~\ref{fig:4}) ranging from 3.5~\AA{} to 5.2~\AA{}, weighted by the area of a sphere $4\pi r^2$ times the vacancy number density $\sim$0.00368 /\AA{}$^3$~(Fig.~\ref{fig:5}a).
A vacancy located at a cation lattice site in \mbox{c-GST} has 12 first nearest cation sites at a distance of 3$\sqrt{2}$~\AA{}.
The expectation number of the V-V duos is 2.4 for the vacancy-disordered configurations, while it is 6 for the vacancy fully ordered configurations.
The coordination number of V-V duos gradually increased from 2.44 to 2.70 by annealing~(Fig.~\ref{fig:5}a).

The vacancy-RDF second peak at 6~\AA{} originates from the straight vacancy-Te-vacancy (V-Te-V) trio illustrated in the inset of Fig.~\ref{fig:5}b.
The number of V-Te-V trios is calculated by integrating the vacancy-RDF second peak~(Fig.~\ref{fig:5}b) ranging from 5.3~\AA{} to 6.6~\AA{}.
Since a vacancy has 6 first-nearest Te atoms at a distance of 3~\AA{}, the average coordination number of V-Te-V trios converges to 1.2 for randomly disordered vacancies.
The coordination number of the trios decreased monotonically from 1.2 to 0.2 during the 100~ns annealing process~(Fig.~\ref{fig:5}b).

The increase in the V-V duo coordination by $\sim$11\% and the reduction of V-Te-V trio coordination by $\sim$83\% serve as evidence of vacancy ordering towards planar vacancy layers.
For example, when a Te atom neighbors one vacancy in a direction (e.g., \textit{x}-axis), the Te atom tends to simultaneously neighbor other vacancies in the other directions (e.g., \textit{y} and \textit{z}-axis).
On the other hand, a situation in which a Te atom neighbors two vacancies in a single direction on both sides is extremely unfavorable. 

Migration of a vacancy is equivalent to the positional exchange of a vacancy with a neighboring cation.
To determine whether Ge or Sb is more responsible for vacancy migration, we calculated the mean squared displacement~(MSD) from the initial RAG configuration during annealing.
See Supplemental Material for the species-wise MSD along the annealing~\cite{RN35}.
During the first 1~ns, the MSD of Sb atoms increased more rapidly and became larger than the MSD of Ge by 0.5~\AA{}$^2$.
The MSD gap between Ge and Sb remained $\sim$0.5~\AA{}$^2$ after 1~ns, indicating that Ge and Sb atoms equally contributed to vacancy migration.
Despite having a smaller covalent radius~(1.21~\AA{}) and lighter mass~(72.64~a.u.) than Sb (covalent radius of 1.40~\AA{} and mass of 121.76~a.u.), Ge did not contribute more to vacancy migration than Sb.

\subsection{\label{sec:3-2}Debye-Waller thermal \textit{B} factor}

\begin{figure}[t]
\includegraphics[width=1.0\linewidth]{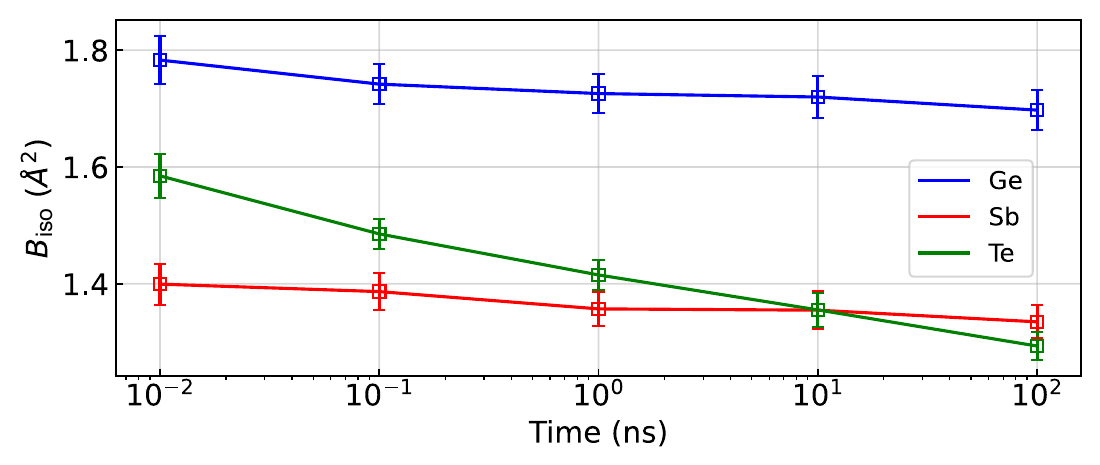}
\caption{\label{fig:6}
Change in $B_\textrm{iso}$(\textit{T}=300~K) by annealing.
The error bars indicate the standard deviation of the thermal factors calculated from equilibrium MD simulations.}
\end{figure}

The Debye-Waller thermal isotopic factor $B_\textrm{iso}$$=$$\frac{8\pi^2}{3} \langle u^2 \rangle$ was calculated at five time points (10~ps, 100~ps, 1~ns, 10~ns, and 100~ns), where $\langle u^2 \rangle$ is the mean squared displacement of ions~(Fig.~\ref{fig:6}).
We obtained $B_\textrm{iso}$ at each time point with a 100~ps-long, 300~K, NVT-MD simulation followed by a 50~ps-long thermal equilibration.
The mean thermal factors of Ge, Sb, and Te decreased monotonically until 100~ns.
Notably, the mean thermal factor of Te atoms $B_\textrm{iso}^\textrm{Te}$ decreased significantly by $\sim$19\% (from $B_\textrm{iso}^\textrm{Te}$=1.58~\AA{}$^2$ at 10~ps to $B_\textrm{iso}^\textrm{Te}$=1.29~\AA{}$^2$ at 100~ns).
The vacancy disorder in the initial RAG configuration seem responsible for the abnormally large thermal factor of Te atoms.
The thermal factors of \mbox{c-GST} measured by neutron and X-ray diffraction at room temperature were $B_\textrm{iso}^\textrm{Ge}$=1.63~\AA{}$^2$, $B_\textrm{iso}^\textrm{Sb}$=1.64~\AA{}$^2$, and $B_\textrm{iso}^\textrm{Te}$=1.671~\AA{}$^2$~\cite{RN30}.
The mean thermal factor $B_\textrm{iso}$$\sim$1.58~\AA{}$^2$ of the configuration annealed for 10~ps is close to the measurement.

The sensitivity of $B_\textrm{iso}^\textrm{Te}$ to annealing suggests that the vacancy distribution significantly affects the dynamics of Te lattices.
This result is convincing from our finding that vacancies are always surrounded by the nearby Te atoms.
It also implies that the thermal \textit{B} factor of Te atoms could potentially serve as an indicator of the degree of vacancy ordering.

\subsection{\label{sec:3-3}Boson peak}

\begin{figure}[t]
\includegraphics[width=1.0\linewidth]{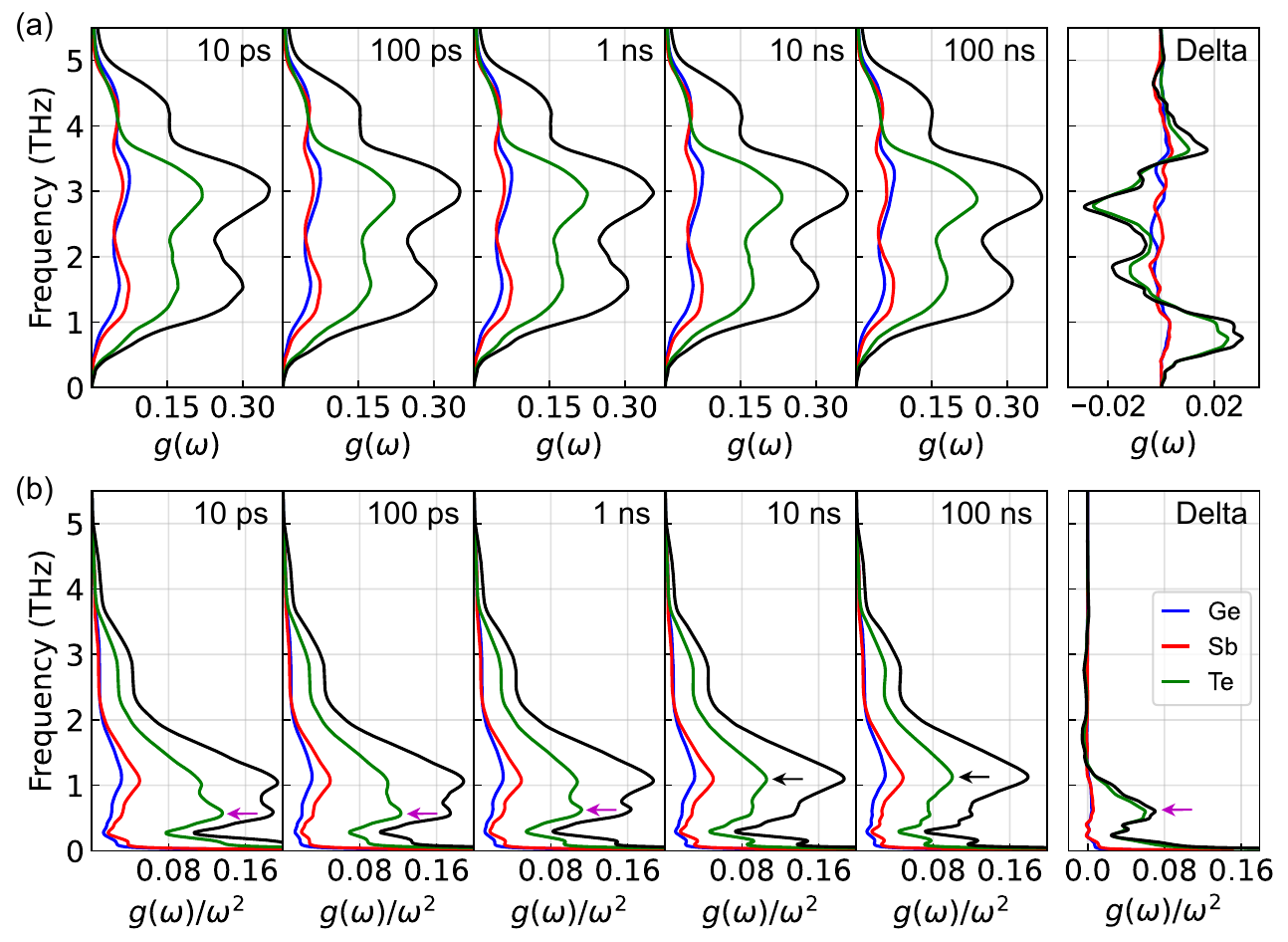}
\caption{\label{fig:7} 
Partial VDOS of \mbox{c-GST} thermally annealed for 10~ps, 100~ps, 1~ns, 10~ns, and 100~ns.
The reduced VDOS in~(b) is scaled by $\omega^{-2}$ from the VDOS in~(a).
The rightmost panel labeled “Delta” shows the subtraction of VDOS at 10~ps by that at 100~ns.
The arrows in purple highlight the distinct boson peaks and those in black the Brillouin peaks.
All curves are smeared by Gaussians with a 0.05~THz width.}
\end{figure}

We calculated the VDOS of \mbox{c-GST} to analyze its dynamic states through vacancy ordering~(Fig.~\ref{fig:7}).
The VDOSs were calculated from the VACF of the same 300~K NVT-MD simulations used to calculate the thermal factors.
The change in the overall shape of the VDOS is negligible, except for the localized vibration modes at $\sim$0.575~THz.
These localized modes are more clearly recognized in the reduced VDOS as scaled by $\omega^{-2}$~(Fig.~\ref{fig:7}b), forming a so-called boson peak.
The VDOS of the configurations at 10~ps, 100~ps, and 1~ns show distinctive boson peaks, while those at 10~ns and 100~ns do not.
Notably, the boson peak modes involve with the vibration of Te atoms. 
Another peak at $\sim$1.1~THz~(Fig.~\ref{fig:7}b) is the Van Hove singularity developed by the TA phonon modes at the zone boundaries, namely the Brillouin peak.
We found that this Brillouin peak is evenly contributed by all elements, indicating the delocalized feature of the phonon modes.
A slight shift in the position of the Brillouin peak is an artifact due to the tail of the nearby boson peak.
The shape of the Brillouin peak does not change as the vacancies become ordered.

\begin{figure}[t]
\includegraphics[width=1.0\linewidth]{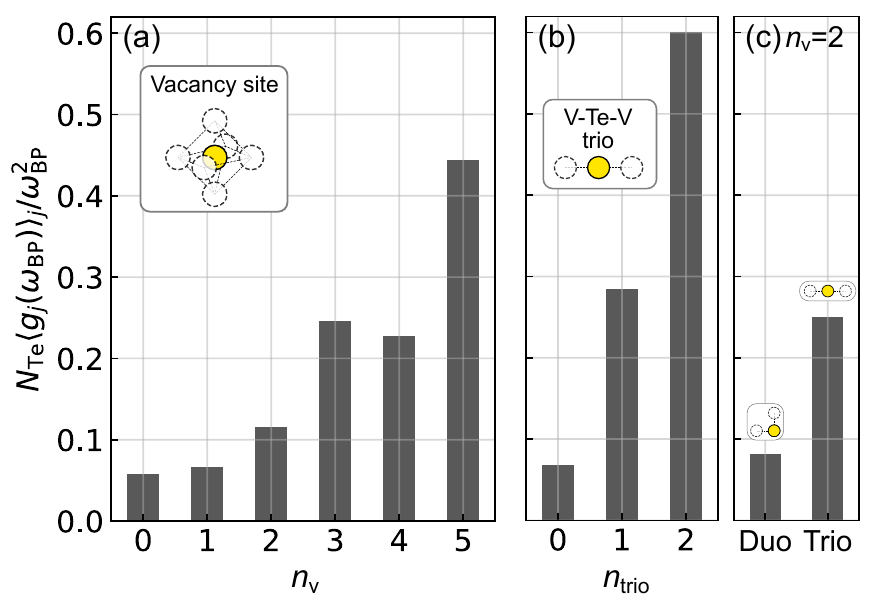}
\caption{\label{fig:8}
Correlation between the vacancy distribution and boson peak intensity.
Configuration-averaged boson peak intensity projected into a Te atom in terms of (a) the number of vacancies ($n_{\rm{v}}$) at the cation sites nearest to Te, (b) the number of straight trios ($n_{\rm{trio}}$), and (c) the two characteristic configurations (V-V duo and V-Te-V trio) of Te atoms with two neighboring vacancies ($n_{\rm{v}}$=2).
The boson peak intensity is scaled by the number of Te atoms $N_{\rm{Te}}$(=6912) for comparison to the boson peak in Fig.~\ref{fig:7}.
}
\end{figure}

The normal mode analysis for the boson peak vibrational modes showed that the boson peak consists primarily of the vibrational modes localized at Te atoms (See Supplemental Material for visualization of the atom-wise boson peak intensity~\cite{RN35}).
The boson peak intensity projected on atom \textit{j} provides a strong signature of the correlation between the vacancy distribution and the boson peak.
We calculated the atom-projected boson peak intensity $g_j(\omega_{\rm{BP}})$$/$$\omega_{\rm{BP}}^2$ from the atom-projected VDOS $g_j(\omega)=\frac{1}{3N}\sum_{i}{\delta(\omega-\omega_i)\sum_{\alpha}{u_{i,j\alpha}^2}}$, where $u_i$ is the polarization vector of the \textit{i}-th vibrational normal mode, \textit{N} the total number of atoms, and $\alpha$= \textit{x}, \textit{y}, or \textit{z}.
We found that configuration-averaged value of the atom-projected boson peak intensity $g_j(\omega_{\rm{BP}})$$/$$\omega_{\rm{BP}}^2$ is proportional to the number of vacancies at the cation sites nearest to Te atom \textit{j} (Fig.~\ref{fig:8}a).
Moreover, the boson peak is contributed most by Te atoms that have a specific arrangement of neighboring vacancies.
In Fig.~\ref{fig:8}b, we observe that the boson peak intensity is almost linear to the number of V-Te-V trios.
This finding becomes more apparent when we compare the contribution to the boson peak from the modes localized at Te atoms with two neighboring vacancies.
The contribution is large when a Te atom and two vacancies form a straight trio (V-Te-V) but, for V-V duos, it is almost the same as the case for 0 or 1 neighboring vacancy (Fig.~\ref{fig:8}c).
We note that the V-V duo formation is the building block for the vacancy layers (i.e., vacancy ordering) as found in the hexagonal structure.

These findings support that the disappearance of the boson peak is directly attributed to the vacancy ordering and, in other words, that the structural disorder in \mbox{c-GST} contributes to the development of the boson peak.
The boson peak intensity in \mbox{c-GST} appears to be highly correlated with the large thermal factor of Te atoms.
A strong correlation between boson peak modes and thermal factors was previously observed in a two-dimensional power-law potential model~\cite{RN31}.

\section{\label{sec:5} Discussion}

We constructed a machine-learned GST225 potential using the RAG dataset and simulated 100~ns annealing of \mbox{c-GST} at 700~K.
After annealing, the initial RAG configuration of \mbox{c-GST} with vacancy disorder transformed into a vacancy semi-ordered cubic structure with a lowered energy by $\sim$20 meV/atom.
The structural transition was captured by the \mbox{\textit{p}-bonding} chain model and the vacancy distribution analysis.
In the final vacancy semi-ordered configuration, Sb atoms occupied the terminal-cation sites in the \mbox{\textit{p}-bonding} chains 21\% more frequently than Ge atoms.
The number of \mbox{\textit{p}-bonding} chains longer than 16 decreased by annealing, while the number of shorter \mbox{\textit{p}-bonding} chains increased.
The change in vacancy distribution also supports vacancy ordering.
Vacancies tend to be located next to Te atoms already neighboring other vacancies, while the configuration of a straight V-Te-V trio is extremely unfavored.
Our annealing simulation of 100~ns may not be long enough to achieve vacancy ordering on a full scale, but all observed structural features consistently indicate the progression of layer-like vacancy ordering.

We identified several key features accompanying the vacancy ordering.
A boson peak develops at $\sim$0.575~THz due to the vacancy disorder at the initial stages of annealing~($t<$1~ns) and then gradually disappears as vacancies are ordered.
This formation and evolution of the boson peak by the vacancy ordering is in sheer contrast to the common observation that the boson peaks are formed in amorphous materials.
The abnormal boson peak in \mbox{c-GST} is attributed to the modes at Te atoms with certain but energetically unfavorable configurations, while the lattice structure retains the cubic symmetry.
We note that such a dynamical phase transition without a change in the density or symmetry of the lattice structure is a unique feature of \mbox{c-GST}.
The Debye-Waller thermal \textit{B} factor of Te decreased by $\sim$19\% in the vacancy semi-ordered cubic phase, indicating its correlation with the boson-peak suppression.
The observable response of the thermal factor and the boson peak suggests that they can serve as dynamic indicators of the vacancy ordering.
Controlling the localized boson peak modes by the vacancy ordering can be an efficient tool for manipulating the thermoelectric properties of the chalcogenides.
Our findings by large-scale MLMD simulations provide a microscopic understanding of the vacancy ordering, which can be utilized for developing multilevel memories and neuromorphic devices using phase-change materials.

\section*{Data Availability}
The RAG-dataset-trained Ge$_2$Sb$_2$Te$_5$ potential in LAMMPS pair-potential format, the RAG configuration dataset and other data supporting the findings of this study and the computational tools (simulation script and analysis codes) used in this study are available from the corresponding author upon reasonable request.

\begin{acknowledgments}
This study was supported by the National Research Foundation of Korea (NRF) Grant No. 2022R1A2C1006530 funded by the Korea government (MSIT), and by Samsung Electronics Co., Ltd (Grant number IO201214-08146-01).
Supercomputing resources including technical support was provided by Supercomputing Center, Korea Institute of Science and Technology Information (Contract No. KSC-2022-CRE-0366).
\end{acknowledgments}



%

\begin{figure*}
  \centering
  \includegraphics[width=1\linewidth, page=1]{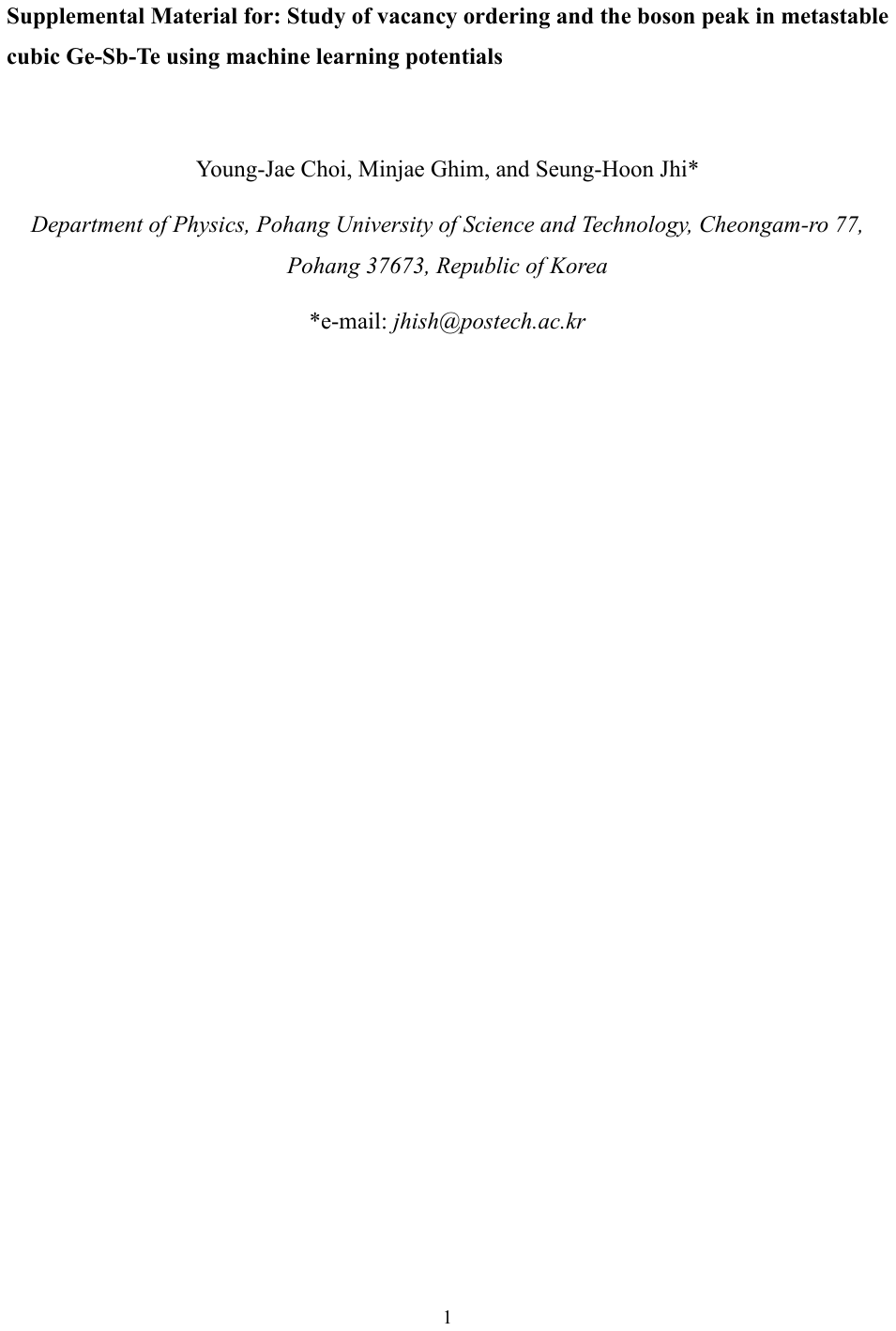}
\end{figure*}

\begin{figure*}
  \centering
  \includegraphics[width=1\linewidth, page=2]{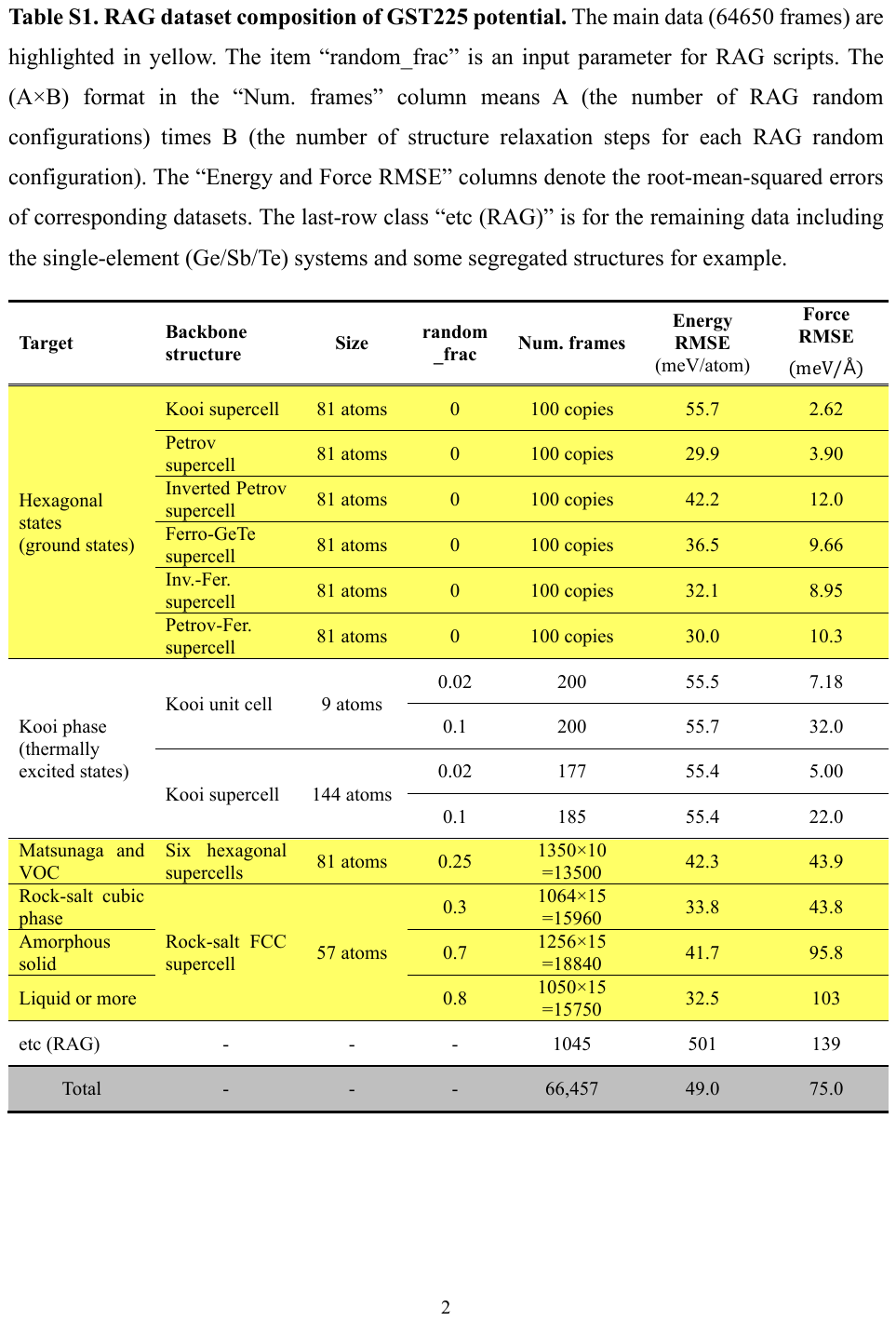}
\end{figure*}

\begin{figure*}
  \centering
  \includegraphics[width=1\linewidth, page=3]{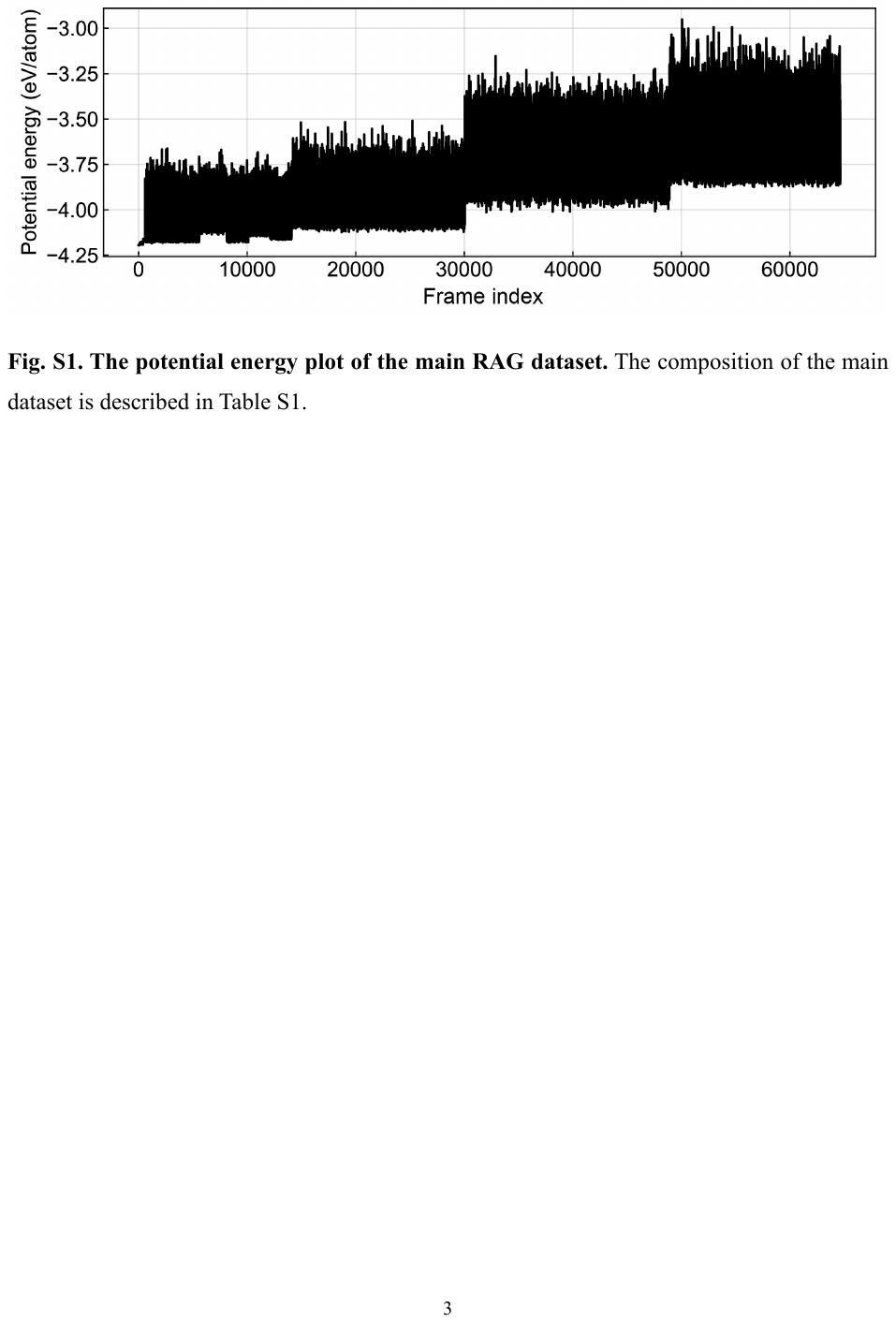}
\end{figure*}

\begin{figure*}
  \centering
  \includegraphics[width=1\linewidth, page=4]{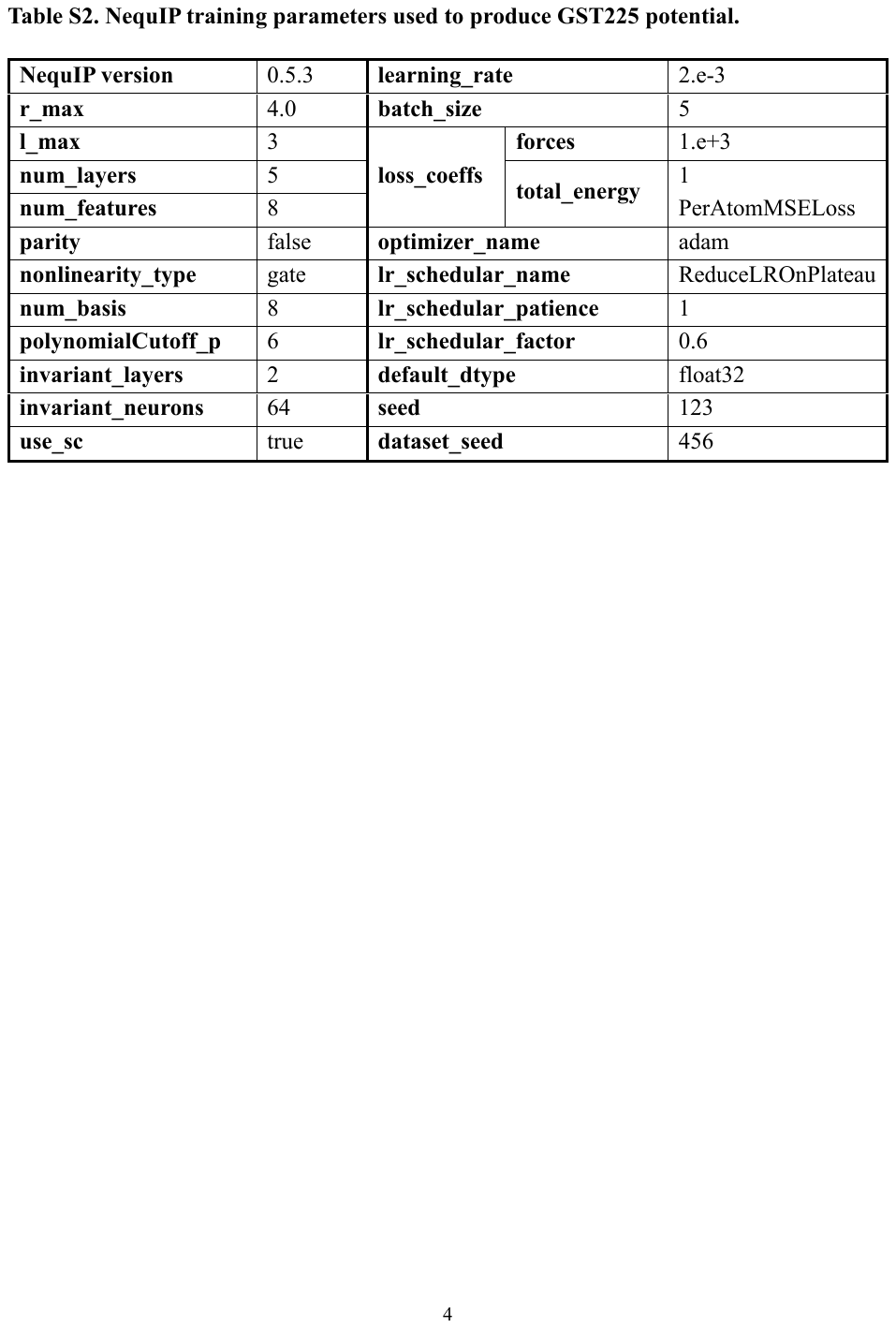}
\end{figure*}

\begin{figure*}
  \centering
  \includegraphics[width=1\linewidth, page=5]{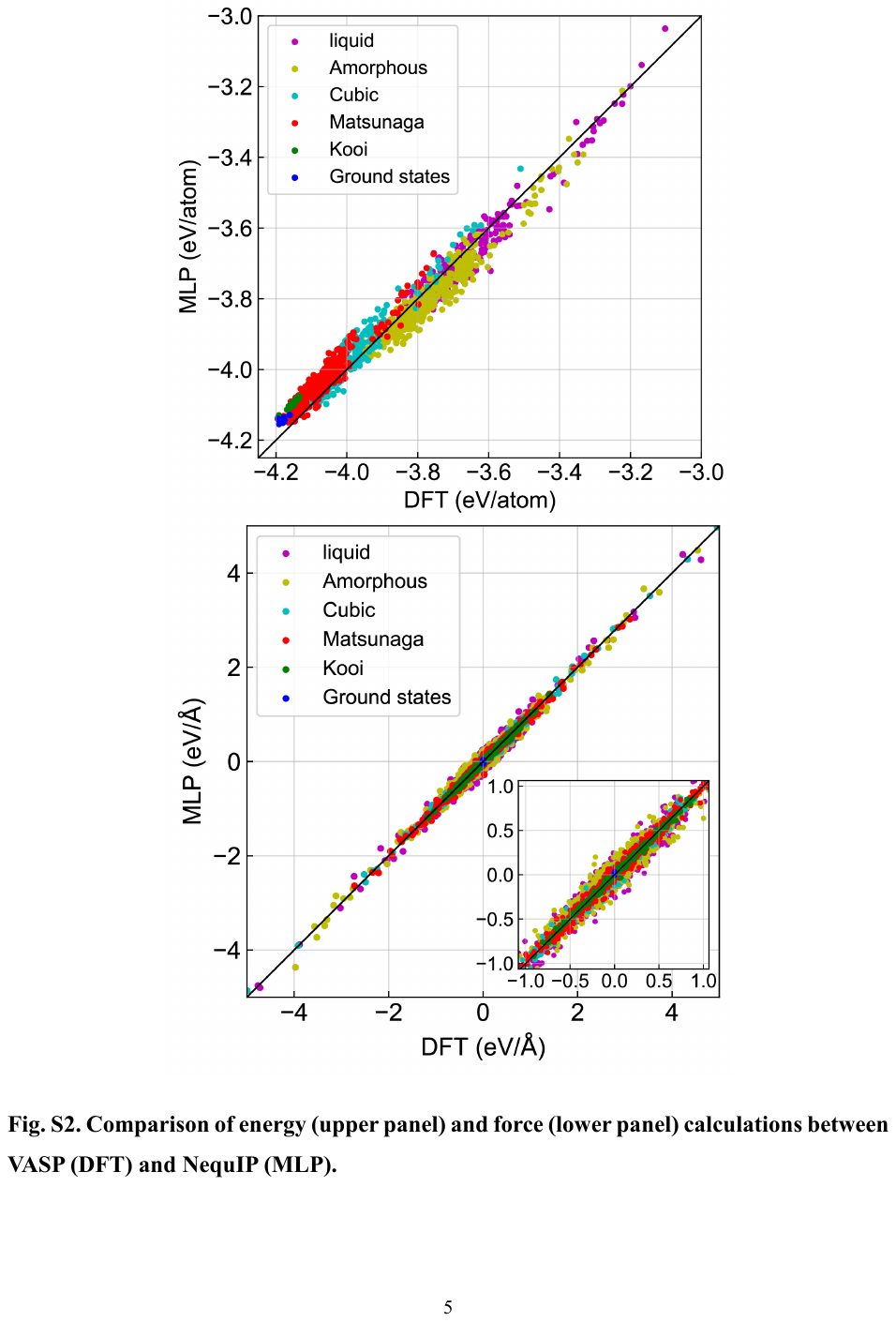}
\end{figure*}

\begin{figure*}
  \centering
  \includegraphics[width=1\linewidth, page=6]{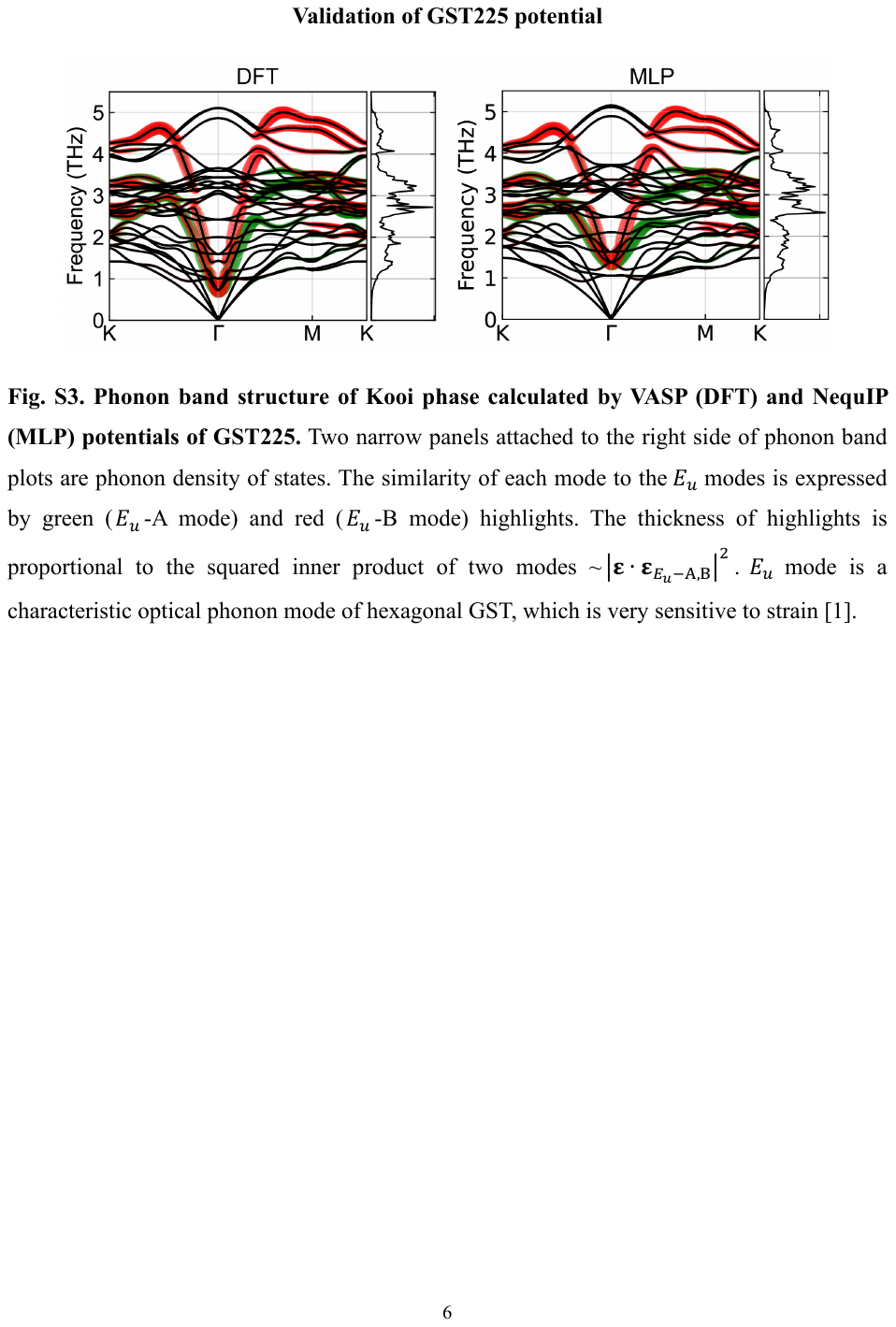}
\end{figure*}

\begin{figure*}
  \centering
  \includegraphics[width=1\linewidth, page=7]{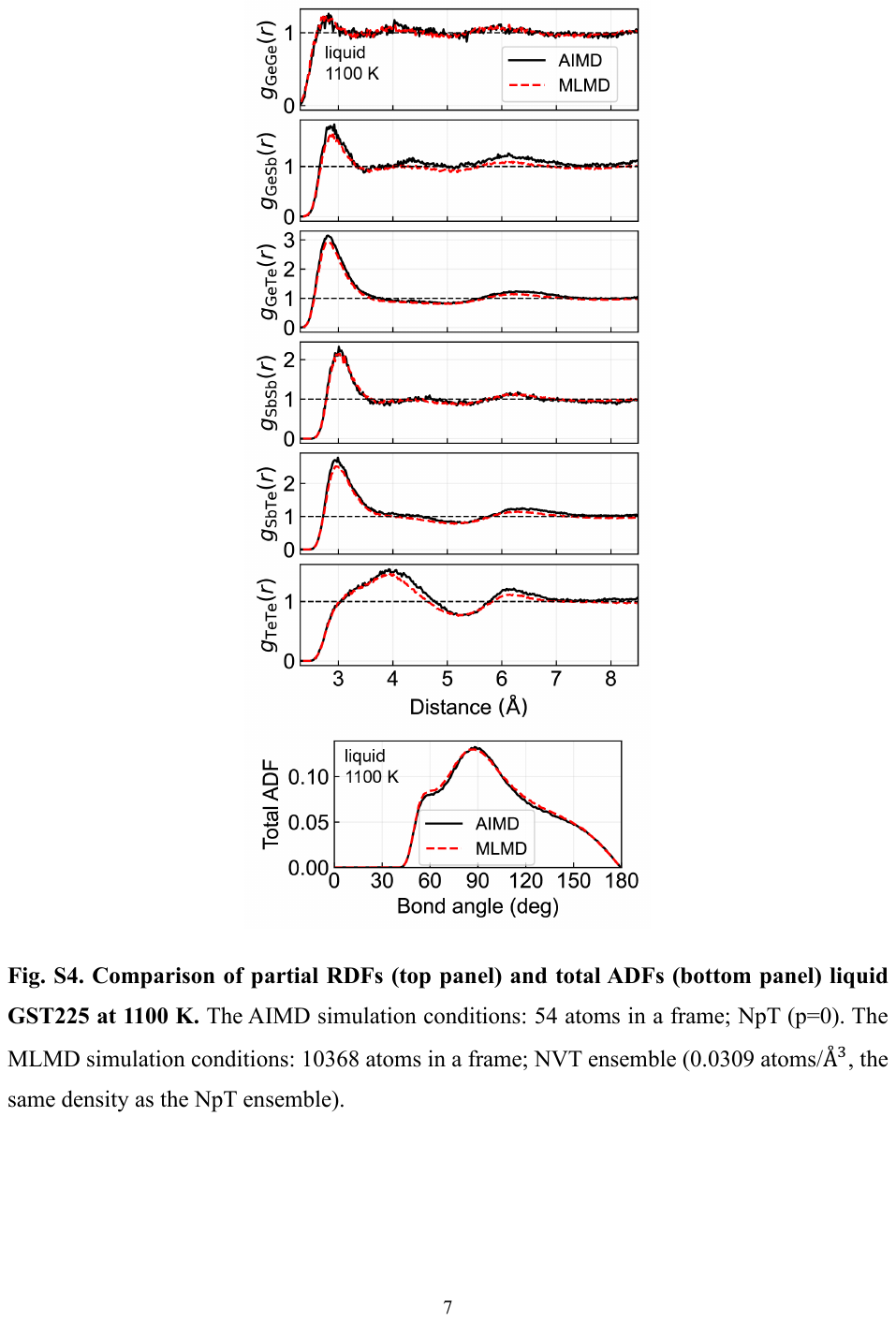}
\end{figure*}

\begin{figure*}
  \centering
  \includegraphics[width=1\linewidth, page=8]{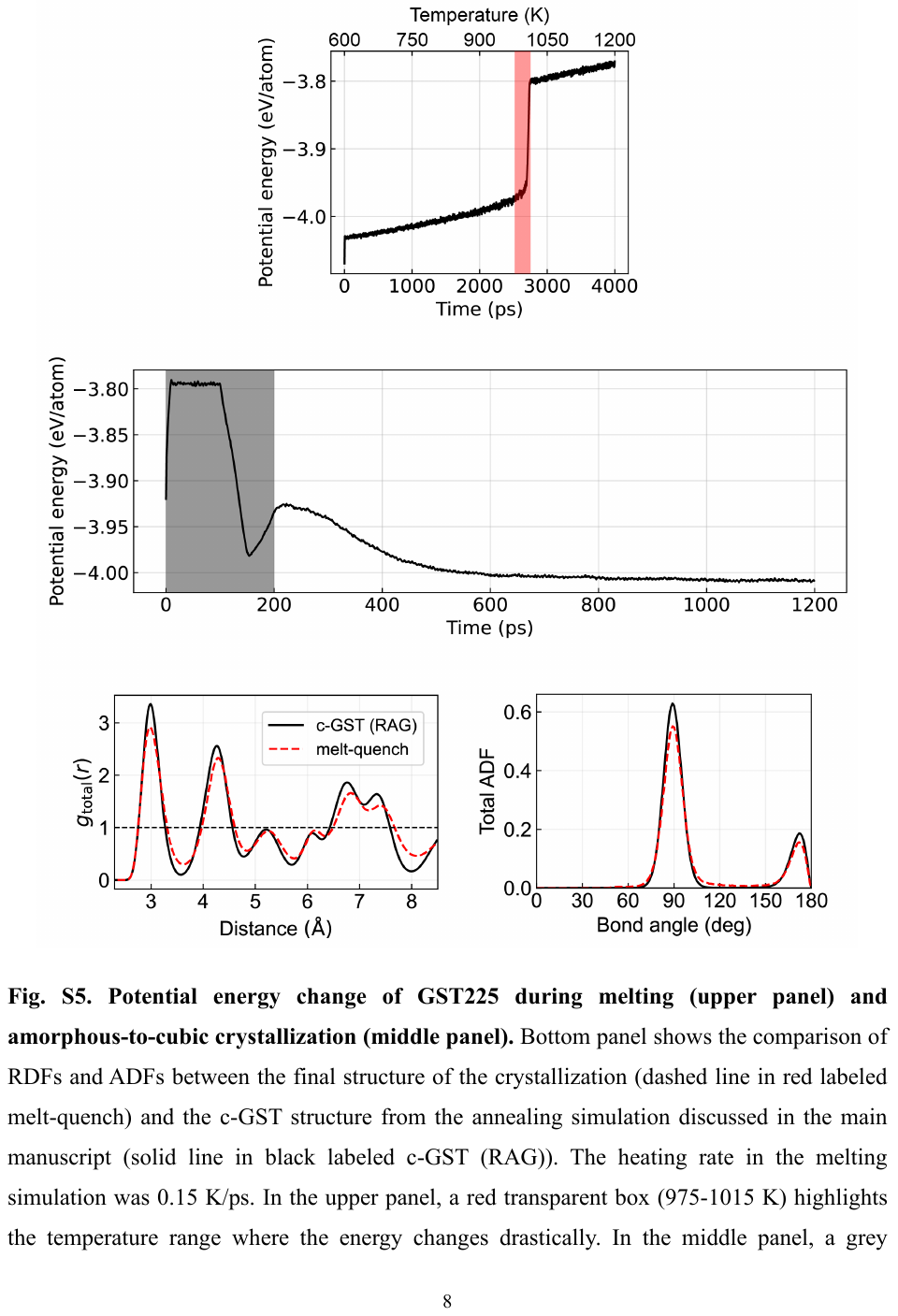}
\end{figure*}

\begin{figure*}
  \centering
  \includegraphics[width=1\linewidth, page=9]{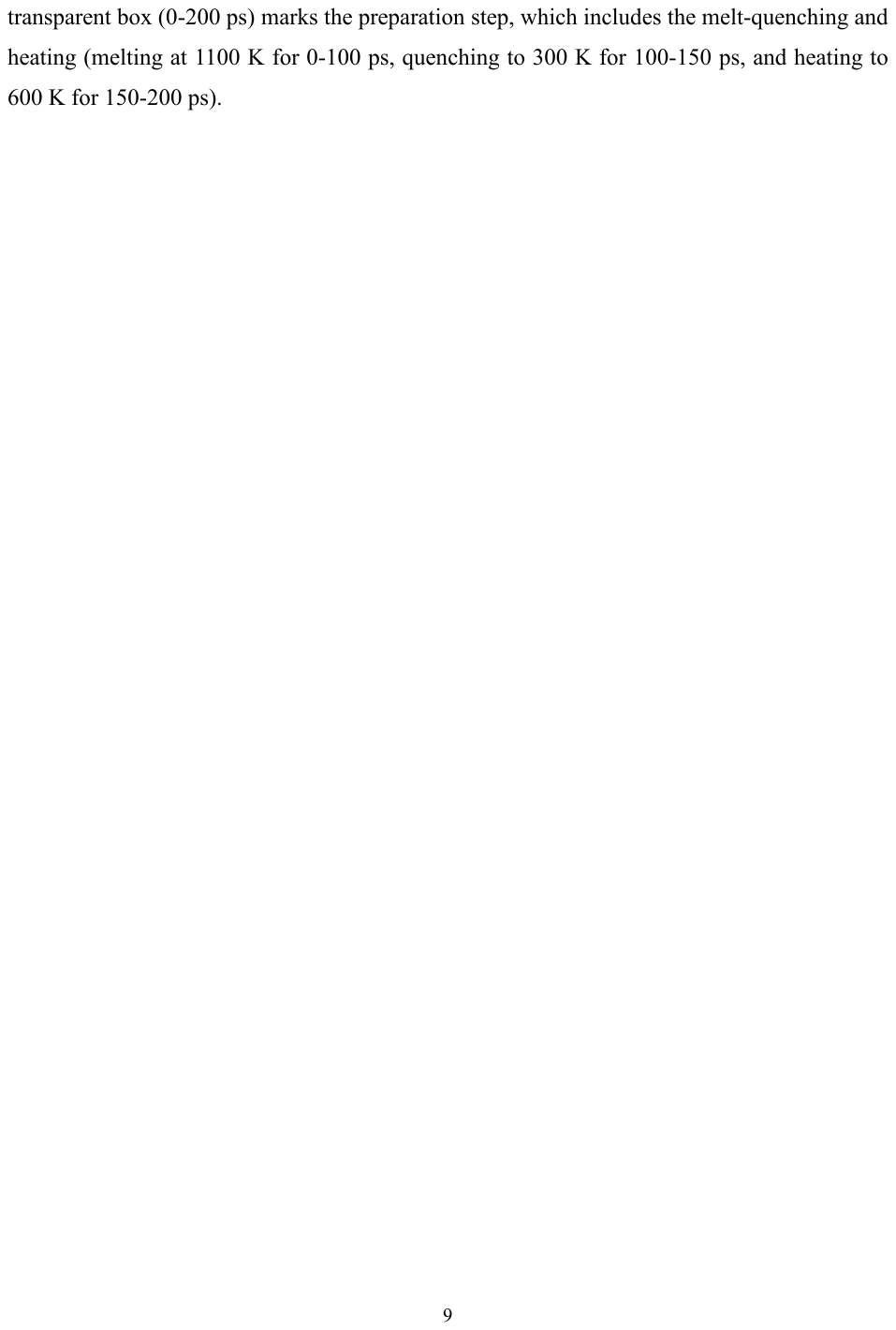}
\end{figure*}

\begin{figure*}
  \centering
  \includegraphics[width=1\linewidth, page=10]{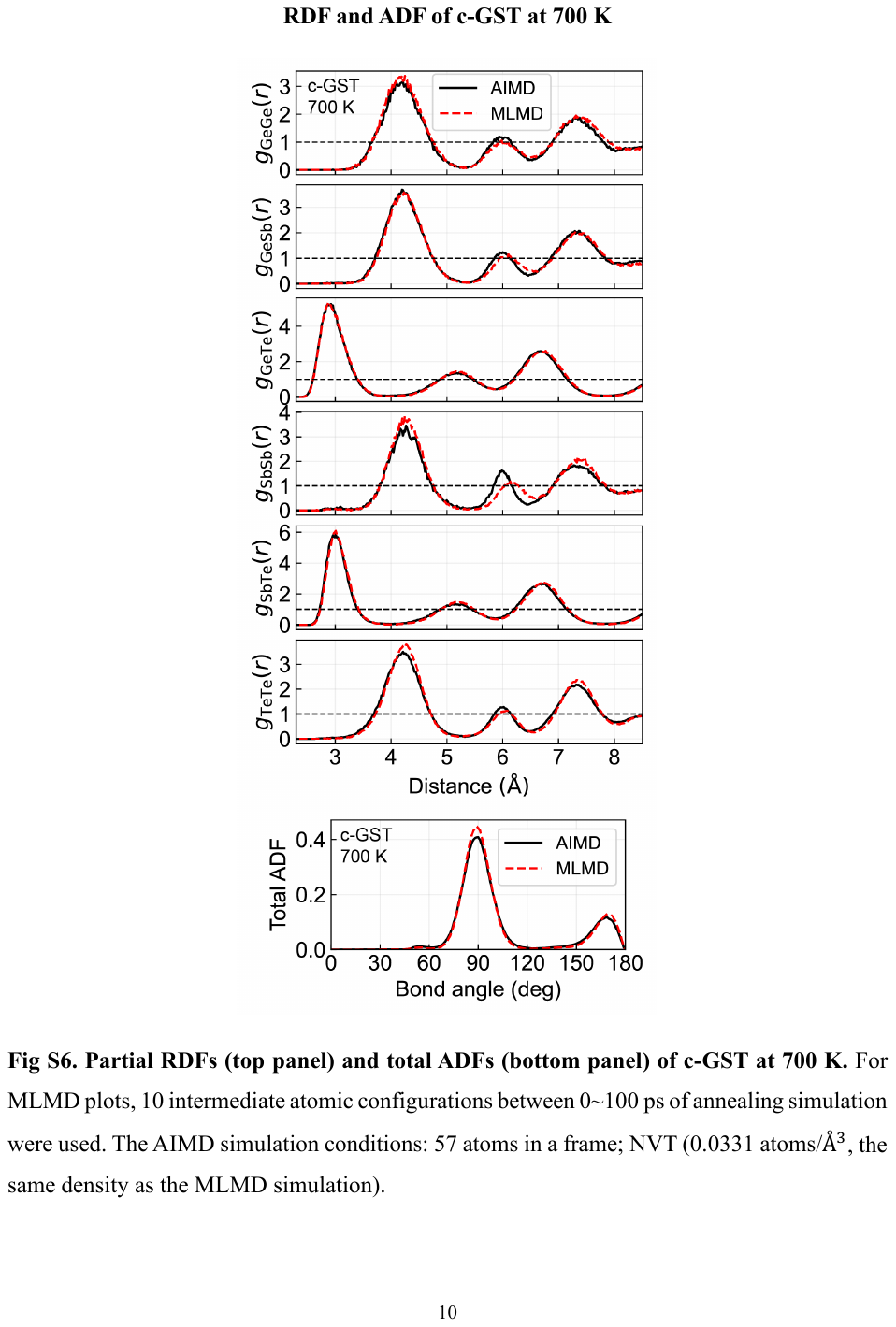}
\end{figure*}

\begin{figure*}
  \centering
  \includegraphics[width=1\linewidth, page=11]{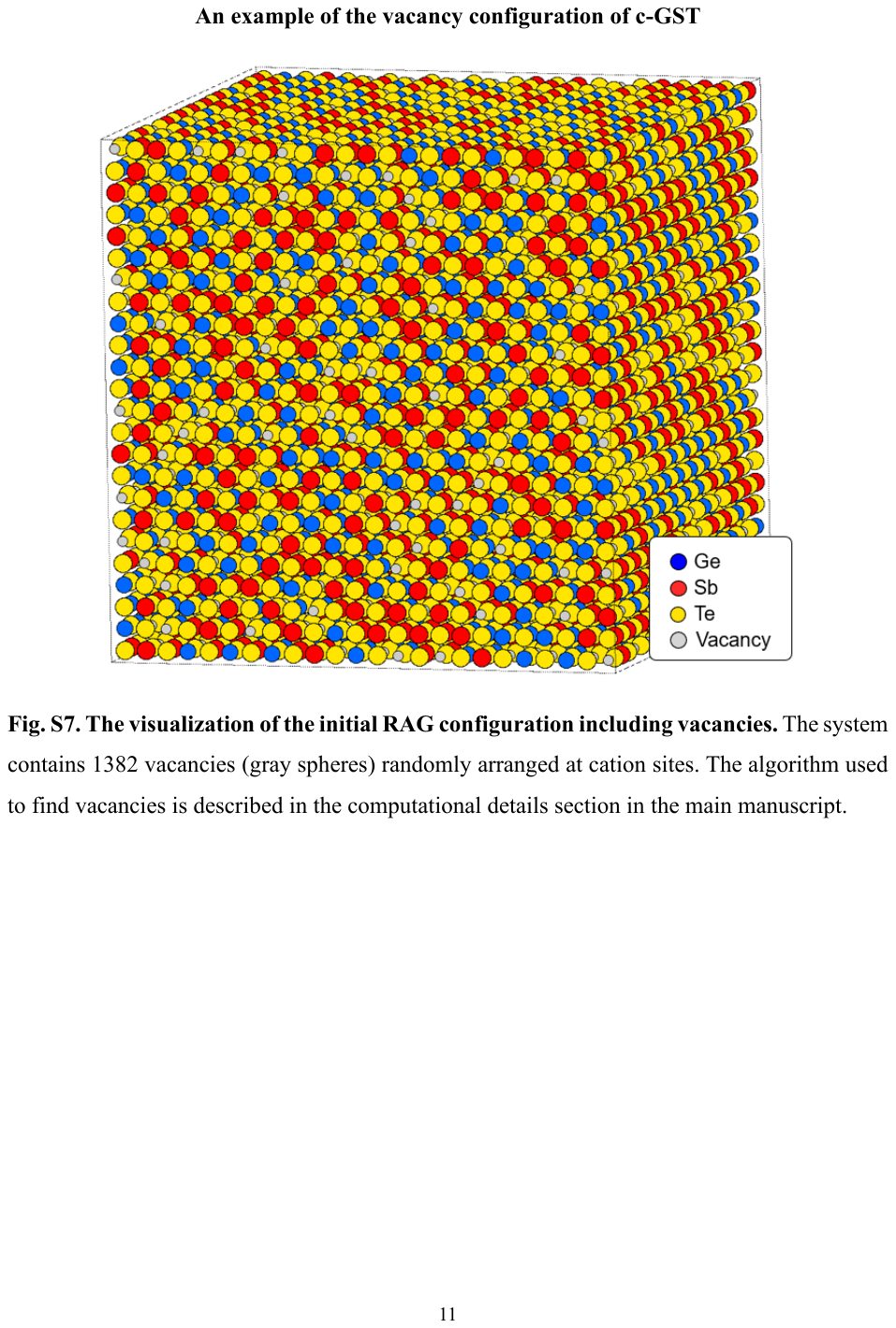}
\end{figure*}

\begin{figure*}
  \centering
  \includegraphics[width=1\linewidth, page=12]{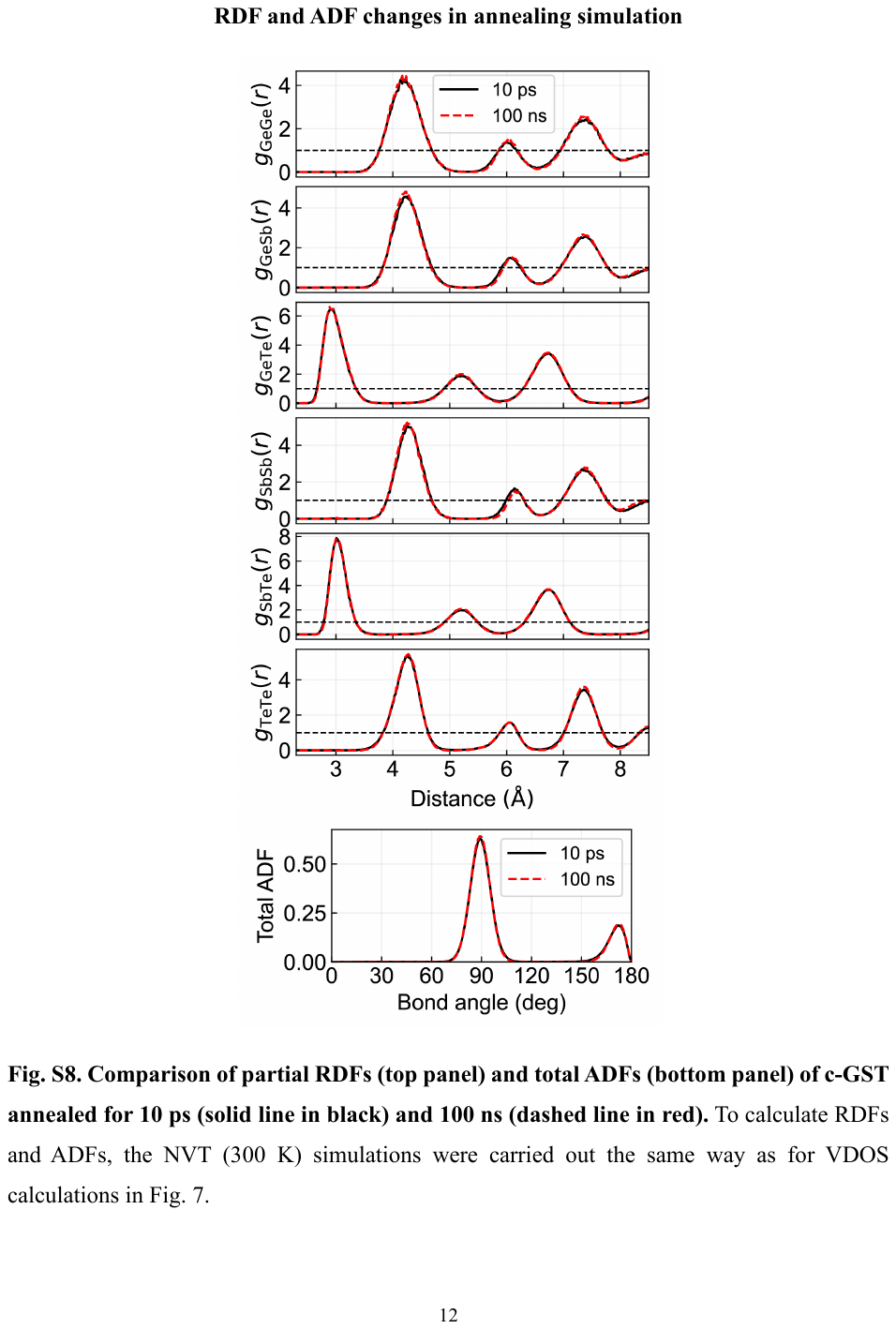}
\end{figure*}

\begin{figure*}
  \centering
  \includegraphics[width=1\linewidth, page=13]{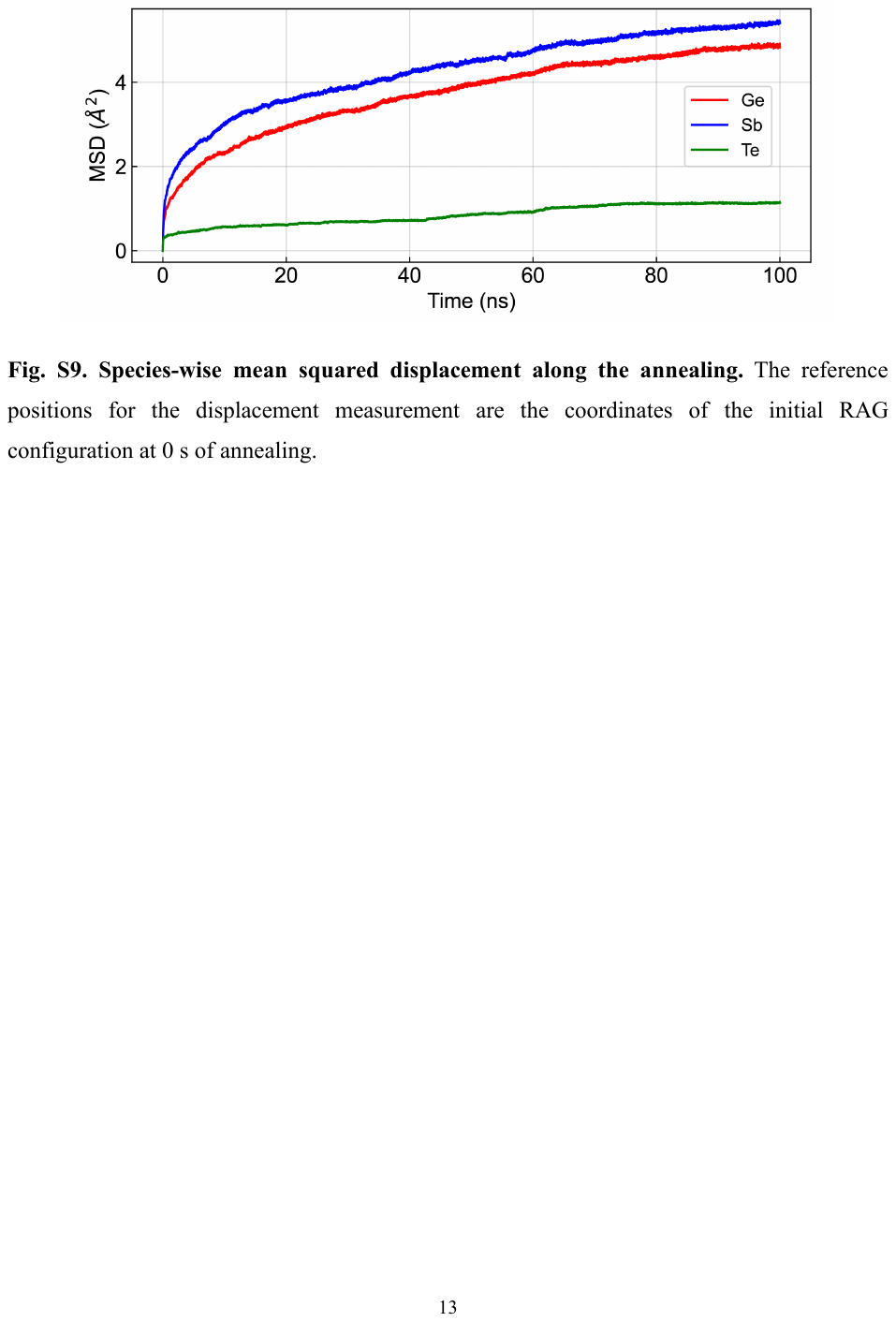}
\end{figure*}

\begin{figure*}
  \centering
  \includegraphics[width=1\linewidth, page=14]{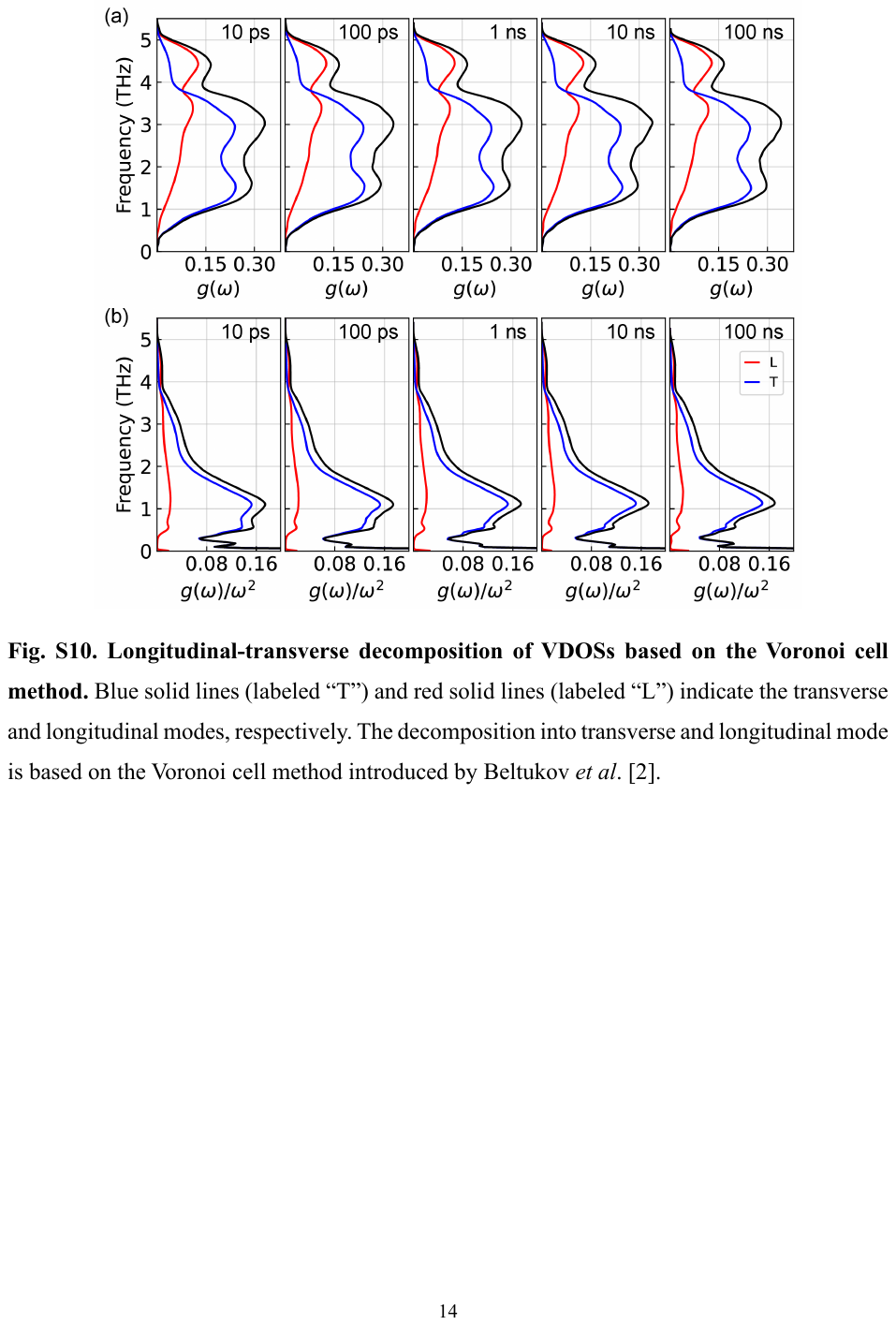}
\end{figure*}

\begin{figure*}
  \centering
  \includegraphics[width=1\linewidth, page=15]{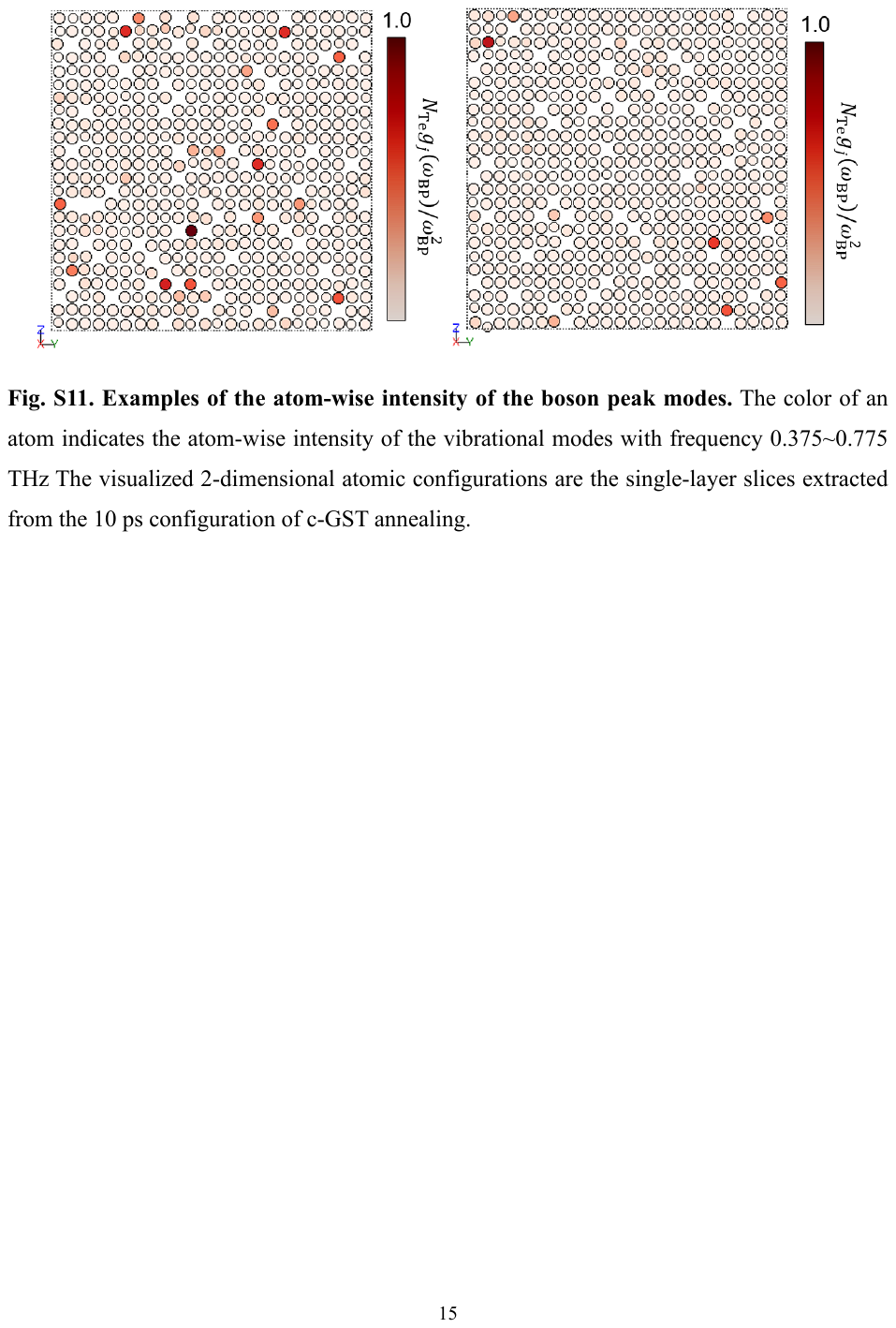}
\end{figure*}

\begin{figure*}
  \centering
  \includegraphics[width=1\linewidth, page=16]{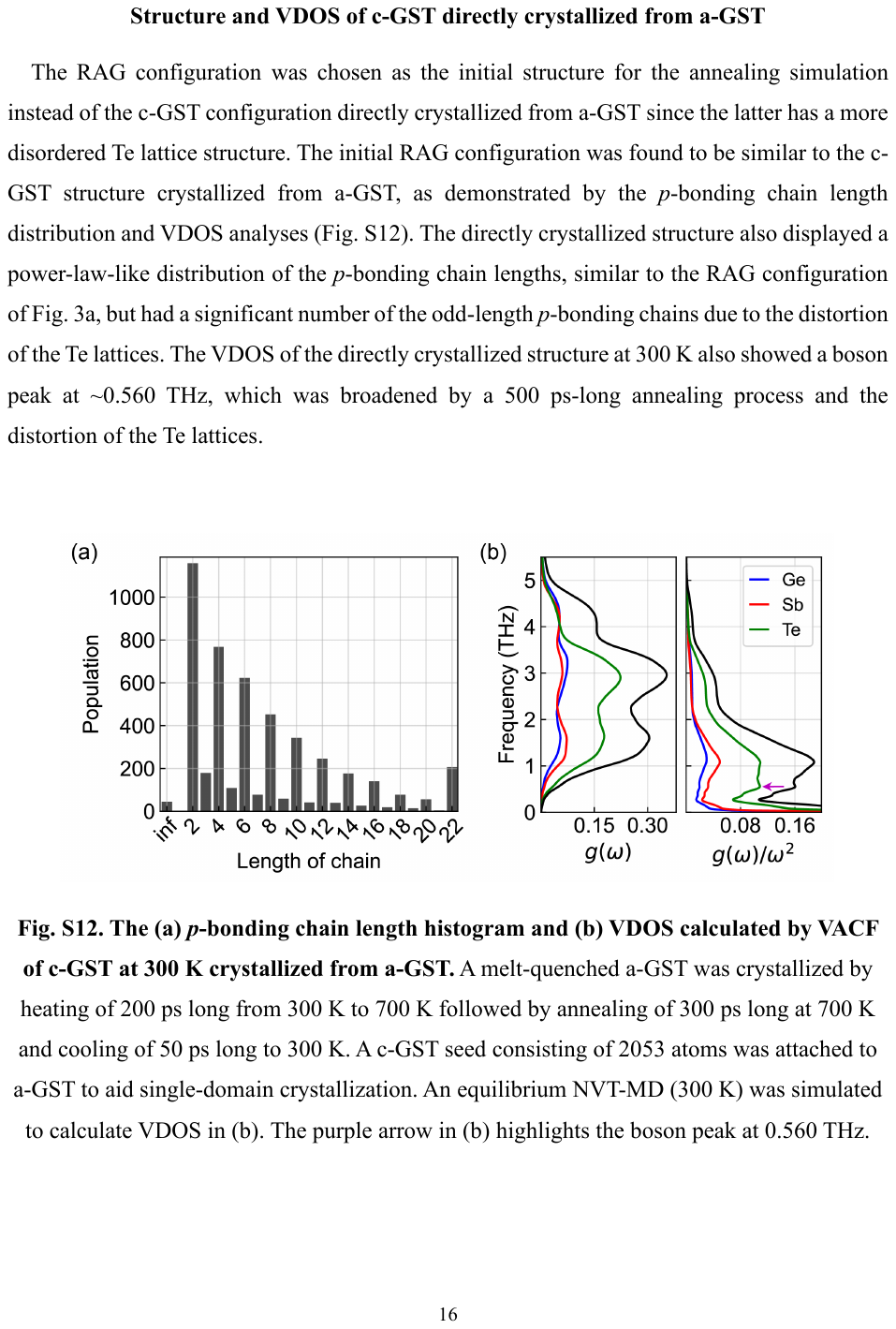}
\end{figure*}

\begin{figure*}
  \centering
  \includegraphics[width=1\linewidth, page=17]{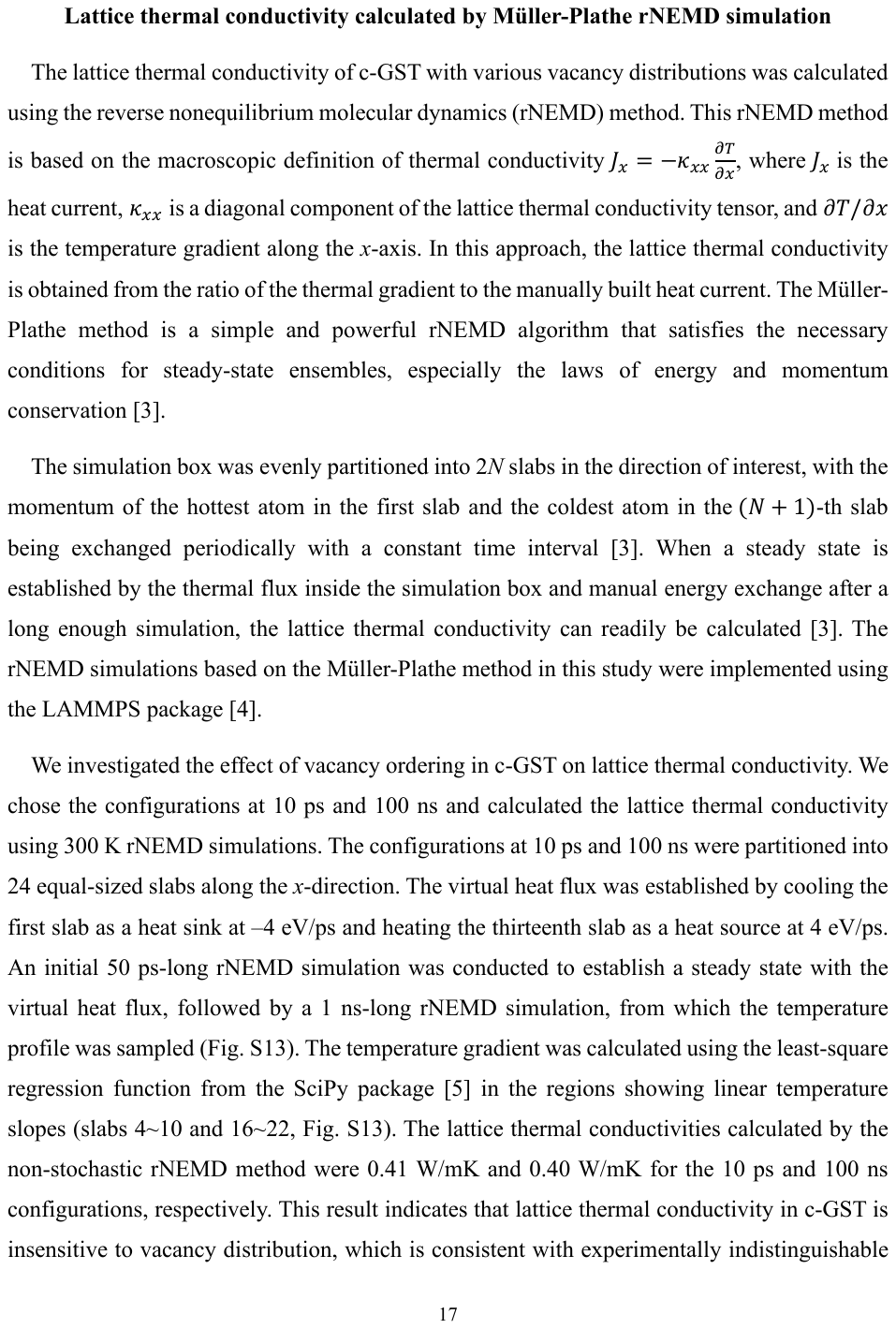}
\end{figure*}

\begin{figure*}
  \centering
  \includegraphics[width=1\linewidth, page=18]{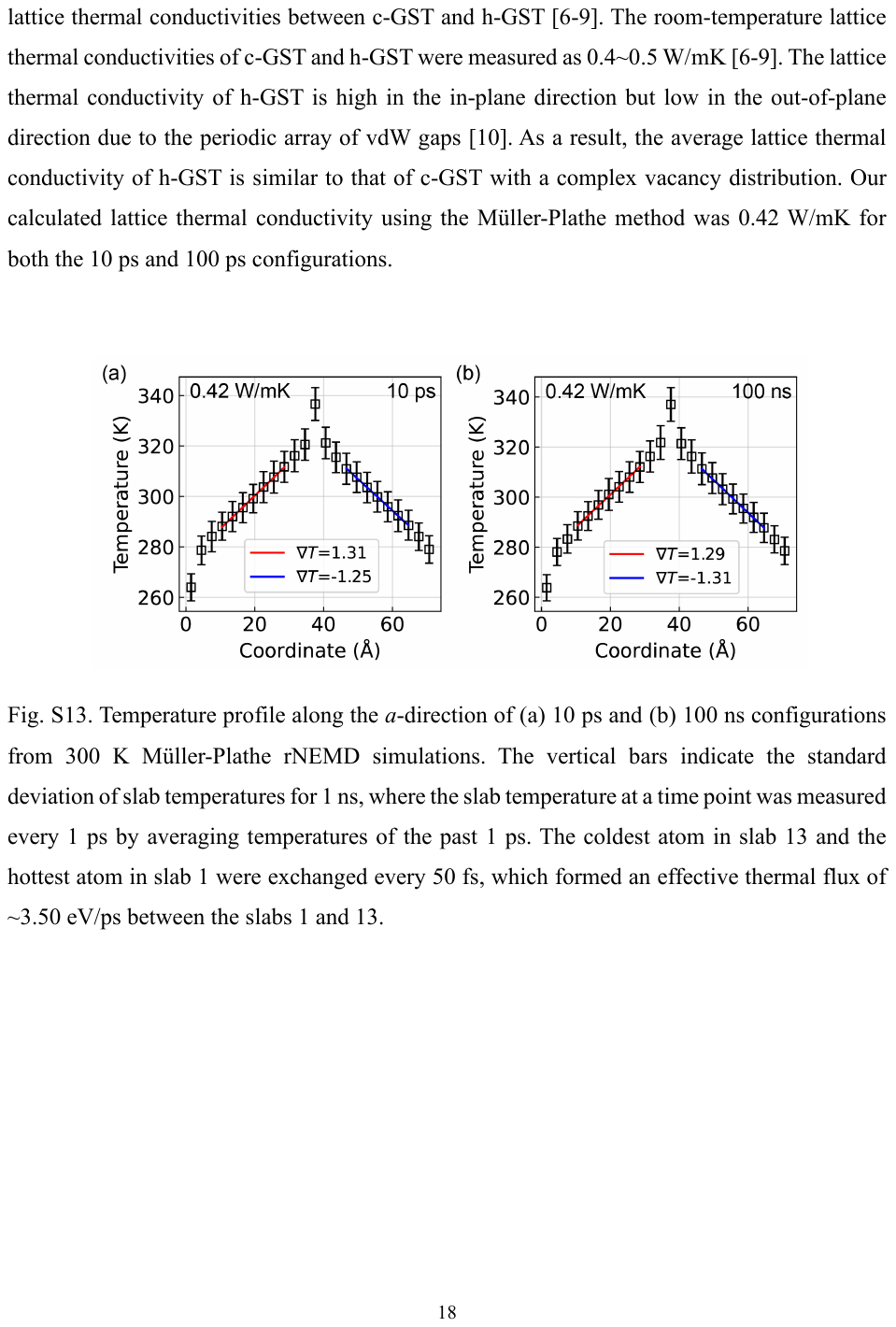}
\end{figure*}

\begin{figure*}
  \centering
  \includegraphics[width=1\linewidth, page=19]{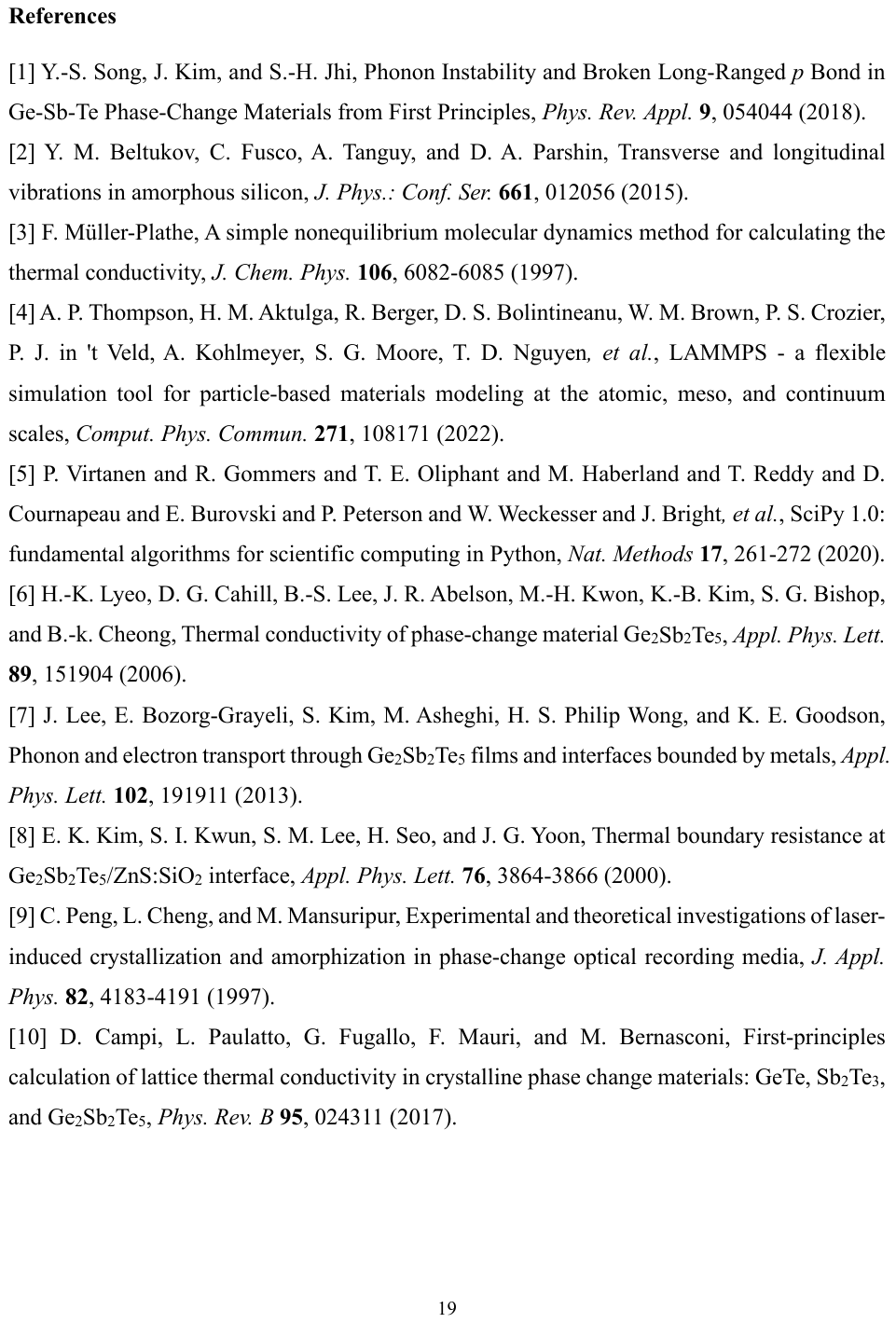}
\end{figure*}

\end{document}